%% file: main.tex
\documentclass{article}

\usepackage{arxiv} 

\usepackage{siunitx} 
\usepackage{graphicx} 
\usepackage{epstopdf}

\usepackage{amsmath} 
\usepackage[utf8]{inputenc}
\usepackage[english]{babel}
\usepackage{comment}

\usepackage[sort&compress,numbers]{natbib}

\newcommand{\vol}{{\ooalign{\hfil$V$\hfil\cr\kern0.08em--\hfil\cr}}}

\renewcommand{\headeright}{ }

\newcommand{\mnflow}{\mbox{\codex1{\textit{mnFlow}}}}

\usepackage[tableposition=top]{caption}
\usepackage{subcaption}
\usepackage{float}
\floatstyle{plaintop}
\restylefloat{table}

\usepackage[utf8]{inputenc}
\usepackage[english]{babel}
\usepackage{hyperref}
\hypersetup{
	colorlinks=true,
	linkcolor=red, 
	anchorcolor=black,
	citecolor=green,
	filecolor=cyan,
	runcolor=cyan,
	menucolor=red,
	urlcolor=magenta, 
}
\usepackage{authblk}

\usepackage{lmodern} 
\usepackage{xcolor} 

\usepackage{soul}
\definecolor{light-gray-x}{gray}{0.9}
\definecolor{pink-x}{rgb}{0.858, 0.188, 0.478}
\definecolor{blue-x}{RGB}{24, 90, 188}
\definecolor{gray-x}{RGB}{241, 243, 244}

\DeclareRobustCommand{\hllgrayx}[1]{{\sethlcolor{gray-x}\hl{#1}}}
\DeclareRobustCommand{\codex1}[1]{{\textbf{\textcolor{blue-x}{\small{\hllgrayx{#1}}}}}}

\usepackage[export]{adjustbox}

\usepackage{listings}
\definecolor{codegreen}{rgb}{0,0.6,0}
\definecolor{codegray}{rgb}{0.5,0.5,0.5}
\definecolor{codepurple}{rgb}{0.58,0,0.82}
\definecolor{backcolour}{rgb}{0.95,0.95,0.92}
\lstdefinestyle{mystyle}{
	backgroundcolor=\color{backcolour},   
	commentstyle=\color{codegreen},
	keywordstyle=\color{magenta},
	numberstyle=\tiny\color{codegray},
	stringstyle=\color{codepurple},
	basicstyle=\ttfamily\footnotesize,
	breakatwhitespace=false,         
	breaklines=true,                 
	captionpos=b,                    
	keepspaces=true,                 
	numbers=left,                    
	numbersep=5pt,                  
	showspaces=false,                
	showstringspaces=false,
	showtabs=false,                  
	tabsize=2
}
\usepackage[ruled,linesnumbered]{algorithm2e}

\SetCommentSty{mycommfont}

\usepackage{textcomp}

\title{Investigation of pressure balance in proximity of 
sidewalls in deterministic lateral displacement}

\author[$\dagger$$\ast$]{Aryan Mehboudi} 
\author[$\dagger$]{Shrawan Singhal} 
\author[$\dagger$$\mathsection$]{S.V. Sreenivasan}

\affil[$\dagger$]{NASCENT Engineering Research Center,
	The University of Texas at Austin, Austin, Texas 78758, United States}
\affil[$\mathsection$]{
	Walker Department of Mechanical Engineering, The University of Texas at Austin, Austin, TX 78712, United States}
\affil[$\ast$]{Corresponding author \textemdash E-mail: aryan.mehboudi@austin.utexas.edu}

\date{March 3, 2025}

\begin{document}
	\clearpage
	\maketitle 
	\thispagestyle{empty}
	
	\input{abstract}
	\newpage
	\input{introduction}

	\input{methods}

	\input{results}

	\input{usage}

	\input{conclusion}
	\input{end_notes}

	
	\bibliographystyle{utphys} 
	\bibliography{dld_bibtex}
	
	

\end{document}

%% file: abstract.tex
\begin{abstract}
Deterministic lateral displacement (DLD) is a popular technique for size-based separation of particles.
One of the challenges in design of DLD chips is to eliminate the disturbance of fluid flow patterns caused by channel sidewalls intersecting with the pillars matrix.
While there are numerous reports in the literature attempting to mitigate this issue by adjusting the gaps between pillars on the sidewalls and the closest ones residing on the bulk grid of DLD array, 
there are only few works that 
also configure the axial gap of pillars adjacent to accumulation sidewall to maintain a desired local pressure field.
In this work, we study various designs numerically to investigate the effects of 
geometrical configurations of sidewalls
on critical diameter and first stream flux fraction variations across channel.
Our results show that
regardless of the model used for boundary gap profile,
applying a pressure balance scheme can improve the separation performance by reducing the critical diameter variations.
In particular, 
we found that for a given boundary gap distribution,
there can be two desired parameter sets with relatively low critical diameter variations. 
One is related to sufficiently low lateral resistance of interface unit cells next to accumulation sidewall, 
while the other one emerges by reducing the axial resistance of the interface unit cells to an appropriate extent.
We believe that this work 
can pave the way for designing DLD systems with improved performance, 
which can be critically important for applications such as 
separation of rare cells, among others,
wherein target species need to be concentrated into as narrow a stream as possible downstream of device to enhance purity and recovery rate simultaneously.

\end{abstract}

%% file: introduction.tex
\section{Introduction}
\label{sec_introduction}

Advancements in microfluidics over the last few decades have created numerous opportunities across industries such as pharmaceuticals, diagnostics, and chemical analysis.
Through the precise control of minute fluid volumes, it allows for the development of portable, low-cost, efficient, and precise platforms for micro/nano-particle manipulation, such as fractionation, sorting, and isolation.
Among many microfluidic separation methods, one may refer to 
magnetophoresis~\cite{munaz_recent_2018}, dielectrophoresis~\cite{pethig_review_2010, zhang_dep-on-a-chip_2019}, and acoustofluidics~\cite{wu_acoustofluidic_2019} as active techniques, while there are various passive microfluidics that can separate species without an external force, such as 
viscoelastic separation~\cite{zhou_viscoelastic_2020},
inertial focusing/separation~\cite{carlo_inertial_2009, zhang_fundamentals_2015},
and various filtration methods.
In contrast with the broad class of physical filtration~\cite{ji_silicon-based_2008}, wherein the cutoff size for particle separation is almost the same as the pore size,
hydrodynamic separation techniques~\cite{huang_continuous_2004,yamada_pinched_2004, takagi_continuous_2005, yamada_hydrodynamic_2005, yang_microfluidic_2006, liang_scaling_2020}
enable fractionation of particles that are smaller than the minimum opening size of the features.

\begin{figure*}[!tb]
\centering
\includegraphics[width=\textwidth]{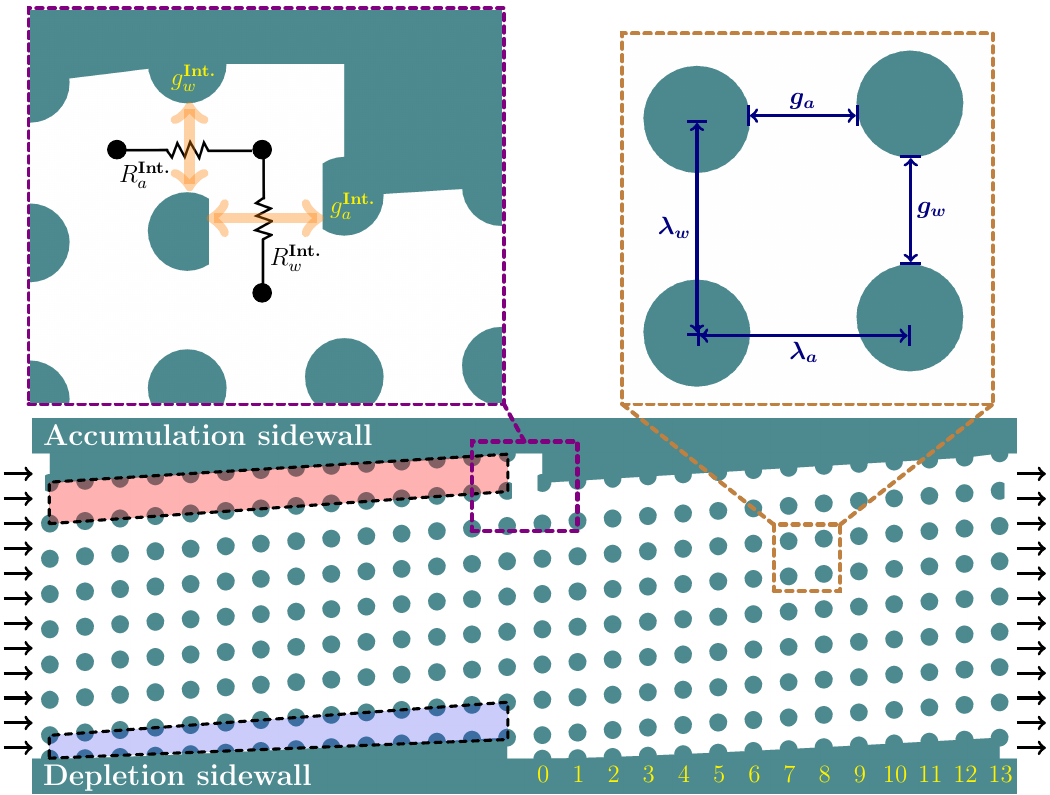}
\caption{Schematic representation of a DLD channel
	of periodicity of $N_p=14$
	with adjustment of lateral gaps of pillars next to the accumulation and depletion sidewalls (filled polygons) and
	with local adjustment of both axial, $g_a^\text{Int.}$, and lateral, $g_w^\text{Int.}$, gaps of pillars next to accumulation sidewall within the 
	interface unit cell, \textit{i.e.}, cell at interface of full DLD units shown in the boxed region next to accumulation sidewall.}
\label{fig_schem}
\end{figure*}

As a member of hydrodynamic separation family,
deterministic lateral displacement (DLD) is a high-resolution separation technique developed by Huang \textit{et. al.}~\cite{huang_continuous_2004}.
It has been widely used for separation of various species such as 
DNA molecules~\cite{wunsch_gel-on-a-chip_2019}, 
exosomes~\cite{wunsch_nanoscale_2016, smith_integrated_2018},
blood components~\cite{davis_deterministic_2006}, etc.
An exemplary DLD structure with two full units periodically repeating along the channel axis is shown in Fig.~\ref{fig_schem}.
Each full DLD unit consists of a matrix of pillars with specific axial ($g_a$ and $\lambda_a$) and lateral ($g_w$ and $\lambda_w$) gap and pitch, respectively.
A working fluid is assumed to flow through the channel from left to right forming $N_w$ fluidic lanes.
The periodicity of DLD ($N_p$) refers to the number or rows of pillars in each unit, and, in conjunction with other geometrical configurations, dictates the critical diameter ($d_c$), \textit{i.e.}, a threshold size particles larger than which follow the tilt angle of pillars matrix ($\tan^{-1} (\lambda_w/N_p\lambda_a)$), while smaller particles show zigzag motion mode following the average fluid flow direction.
The underlying reason for the emergence of these two distinct particle trajectories is that 
fluid flowing through a gap between two adjacent pillars on the same row splits into $N_p$ streams marked out by stagnation streamlines.
The center of a particle with a radius larger than the width of first stream next to a pillar ($\beta$) falls outside of the first fluid flow lane and, despite being in the proximity of pillar, follows the second (or farther) streams.
Therefore, this large particle (ideally) does not penetrate through the axial gaps between pillars, but rather remains in the same fluidic lane, migrating towards the accumulation sidewall and is depleted next to depletion sidewall.
More details can be found in other works, \textit{e.g.}, Salafi \textit{et. al.}~\cite{salafi_review_2019}.

Despite the complex three-dimensional fluid flow dynamics and fluid-wall-particle interactions, there are multiple 
models to estimate critical diameter. 
As a naive model, critical diameter can be approximated as $2\beta$.
The first stream width itself, $\beta$, can be approximated by assuming a Poiseuille flow through axial gap between adjacent pillars residing on the same row~\cite{inglis_critical_2006}.
Considering the experimental data from the literature, 
this model underestimates the critical diameter~\cite{davis_microfluidic_2008}.

One of the challenges of designing DLD structures is to mitigate the fluid flow disturbance in proximity of accumulation and depletion sidewalls, which stems from the sidewalls intersecting with pillars array and blocking their periodic structure in lateral direction.
There are several reports in the literature studying these effects~\cite{inglis_efficient_2009,pariset_anticipating_2017,feng_maximizing_2017,ebadi_efficient_2019,inglis_fluidic_2020}.
In one of the early works, Inglis~\cite{inglis_efficient_2009} showed that in the absence of an appropriate boundary profile,
the capability of DLD in laterally displacing large particles can be impaired significantly.
The work also demonstrated that 
the accumulation and depletion sidewalls can be configured 
according to an analytical model
to reduce the local critical diameter variations and to improve the separation performance.
The developed model was based on the assumption that 
flux of fluid through a gap is proportional to square of gap.
Later, Feng \textit{et. al.}~\cite{feng_maximizing_2017} revised the model by considering 
the flux of fluid through a gap being proportional to cube of gap
analogous to fluid flow between shallow infinitely wide channels. 
As plane Poiseuille flow may not accurately represent the fluid flow dynamics around (circular) DLD pillars, Ebadi \textit{et. al.}~\cite{ebadi_efficient_2019} reported an appealing approach
aiming at
obtaining a correlation between gap size and fluid flux for a DLD unit cell consisting of two semi-pillars.
They conducted two-dimensional numerical simulations for various gap sizes and performed curve fitting by using their curated dataset.
They found a correlation of the form of $Q\propto g_w^b$, wherein $Q$ refers to fluid flux, and $b$ is the fitting parameter reported to be about 2.46, falling between 2.0 
(work of Inglis~\cite{inglis_efficient_2009}) 
and 3.0
(work of Feng \textit{et. al.}~\cite{feng_maximizing_2017}).
The authors have reported an improved fluid flow pattern compared to 
the results they reported for the gap-squared approximation~\cite{inglis_efficient_2009} by reducing the variations of first stream flux.

More recently, Inglis \textit{et. al.}~\cite{inglis_fluidic_2020} have reported a model based on three-dimensional simulation of a DLD unit cell with various aspect ratios.
A significance of the work is that it correlates the design of DLD sidewalls with a larger set of geometrical dimensions of pillars array by including ratios of parameters such as 
pillar height, diameter, as well as axial and lateral pitch.
The work also introduces a model for pressure balance near accumulation sidewall 
by adjusting the lateral and axial gaps of adjacent pillars 
on interface unit cell, \textit{i.e.}, cell at interface of full DLD units as shown in the boxed region next to accumulation sidewall in Fig.~\ref{fig_schem},
to further improve the fluid flow pattern in proximity of sidewall.
The authors reported an improved fluid flow pattern compared to 
the results they reported for the gap-squared approximation~\cite{inglis_efficient_2009} by reducing the variations of first stream flux.

Despite all the aforementioned attempts, a comprehensive study of the pressure balance scheme near accumulation sidewall proposed by Inglis \textit{et. al.}~\cite{inglis_fluidic_2020} (Eq.~1)
and its effects on fluid flow characteristics across channel is still lacking. In particular, there are several open questions that we aim to address as discussed in the following.
\begin{itemize}
	\item For any given DLD system, there can be an infinite number of combinations of axial and lateral resistances that satisfy the pressure balance equation.
	What are the effects of selecting various combinations?
	Is there any optimal pair of axial and lateral resistances?
	Is there a guideline for systematically choosing the desired pair of resistances?
	\item Can the pressure balance scheme be integrated with different boundary treatment approaches discussed above? How would it affect the fluid flow characteristics?
	\item The performance of gap-squared model is usually used as a baseline in the literature, 
	but how can the performance of other available boundary treatment approaches be ranked?
\end{itemize}
In order to address these questions, we design and numerically study various DLD systems in accordance with different sidewall adjustment approaches.

The remainder of this paper is organized as follows. 
The methodology is described in Section~\ref{sec_method}.
In Section~\ref{sec_res}, we present our results and discussion.
We provide a practical guidance in Section~\ref{sec_practical_guide} on how to use the findings of our work in future research.
We conclude in Section~\ref{sec_conclusion} with a brief summary.

%% file: methods.tex
\section{Methods}
\label{sec_method}

\subsection{Numerical simulations}
\label{sec_simulation}

\begin{figure}[!tb]
	\centering
	\includegraphics[width=1\textwidth]{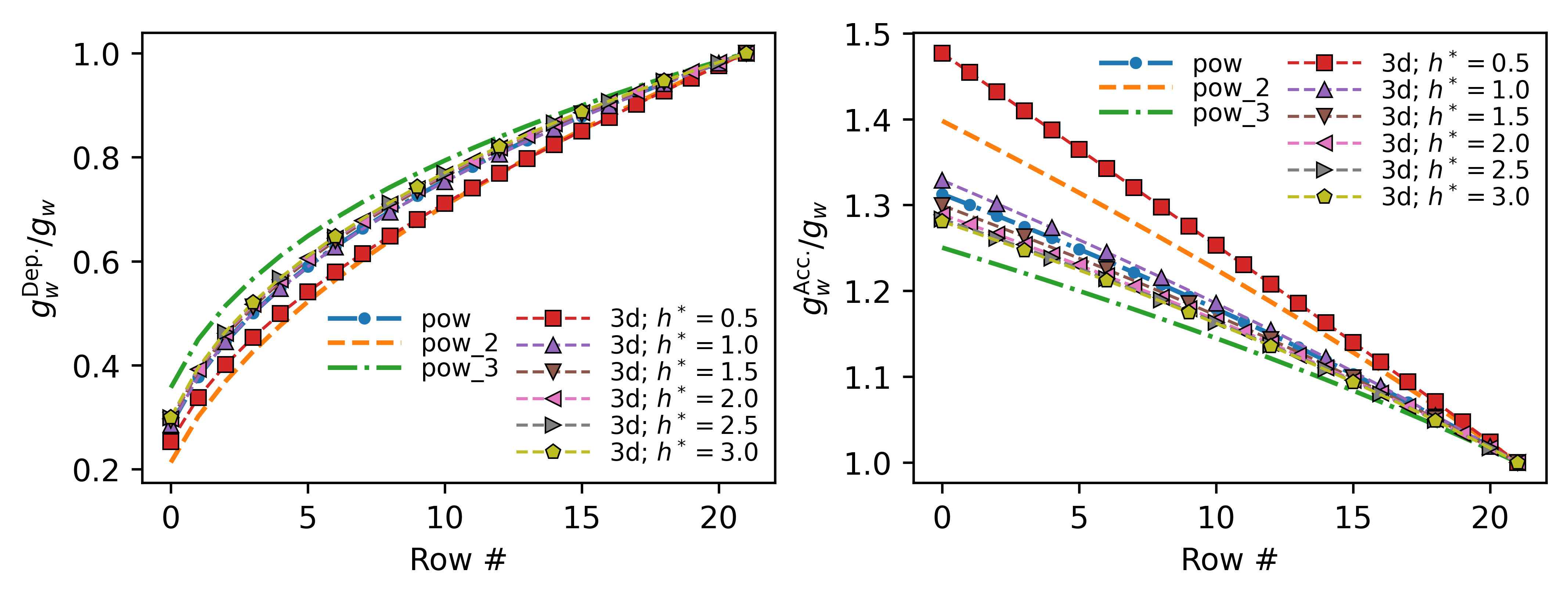}
	\caption{Dimensionless boundary gaps for depletion 
	(left) 
	and accumulation 
	(right) sidewalls
	of a DLD channel with 
	$N_p=22$, 
	determined by using different models: 
	\codex1{pow\_2}~\cite{inglis_efficient_2009}, 
	\codex1{pow\_3}~\cite{feng_maximizing_2017}, 
	\codex1{pow}~\cite{ebadi_efficient_2019},
	and 
	\codex1{3d}~\cite{inglis_fluidic_2020} for various 
	dimensionless channel depths $h^\ast$.
	Note that the pressure balance mechanism is not applied here, which leaves the boundary gap of accumulation sidewall at Row \#$N_p-1$, \textit{i.e.}, Row \#21 here, the same as that in the bulk of array, \textit{i.e.}, $g_w^\text{Int.}/g_w=1$.
	}
	\label{fig_res_boundary_gaps}
\end{figure}
In this work, we assume that the gaps are sufficiently smaller than the pillars height, wherein the problem can be formulated in a two-dimensional (2D) fashion.
To the best of our knowledge, the work of Inglis \textit{et. al.}~\cite{inglis_fluidic_2020} is the only report in the literature taking the pillars height into account to determine the boundary gap profile.
Other models have been developed by using analytical/numerical models assuming a sufficiently small~\cite{inglis_efficient_2009} or large~\cite{feng_maximizing_2017,ebadi_efficient_2019} channel depth.
The boundary gap profiles estimated by using different models are shown in Fig.~\ref{fig_res_boundary_gaps}.
It can be discerned that the gap profiles obtained  
from \codex1{pow\_2} and \codex1{pow\_3} 
bound those estimated by
\codex1{pow} model as well as those from \codex1{3d} model for sufficiently large channel depths, \textit{e.g.}, the cases that $h^\ast\equiv h/\lambda_w$ equals $1.0$ or larger, wherein $h$ denotes the channel depth (pillars height).
One can also perceive that the gap profiles determined by \codex1{3d} model
start to become independent of channel depth as $h^\ast$ increases and approaches $\sim 2.5$.
From the 
\codex1{3d} model's perspective, channels deeper than this threshold can be considered \textit{sufficiently deep} to represent an infinitely-deep channel. 
Thus, unless otherwise stated, we consider $h^\ast=4.0$ when using the \codex1{3d} model in this work.

We used our recently developed DLD design automation (DDA) tool, which is part of the micro/nanoflow ({\mnflow}) package~\cite{mehboudi_universal_2024,mehboudi_mnflow_2024}, to design the DLD systems.
We used Ansys Fluent to solve the 2D steady-state Navier-Stokes equations, simulating laminar flow and obtaining the pressure and velocity fields.
We considered two full DLD units (each with $N_p$ row) along the channel axis and applied periodic boundary condition on inlet and outlet boundaries. 
Then, we performed point-wise particle tracking by using our developed code and determined first stream width ($\beta$), critical diameter ($d_c\approx 2\beta$), first stream flux fraction (FSFF) and its dimensionless value (FSFF\textsuperscript{$\ast$}) nondimensionalized by its ideal counterpart ($1/N_p$).
We investigated the characteristics across both full DLD units and did not find any noticeable discrepancy, which is expected as we apply periodic boundary condition on inlet/outlet boundaries as opposed to other boundary conditions that set velocity and/or pressure on inlet/outlet~\cite{ebadi_efficient_2019}, wherein inlet and outlet effects inevitably contribute into the fluid flow characteristics.
Therefore, we limited the particle tracking simulations only to the first full DLD unit to manage computational costs.
We also conducted mesh study to ensure the characteristics remain independent of mesh size.

Unless otherwise stated, the following configurations were used for our simulations.
The maximum local Reynolds number was about 0.03.
The tracers diameter was 1 nm.
The fluid viscosity was set equal to $1.003\times 10^{-3}$ Pa.s.
We used 256 tracers per gaps.
For the case of having a gap of $7~\mu m$, therefore, the uncertainty around the calculated first stream width ($\sigma$) should be about $\pm 13.7$ nm.

We used unstructured quad-dominant mesh, \textit{i.e.}, the majority of elements were quadrilaterals, while there was also a relatively small fraction of triangular elements.
We used a relatively conservative set of convergence criteria in our fluid flow simulations, ensuring that the normalized residuals dropped to $10^{-12}$.
The mesh convergence analysis was performed on
a DLD system similar to that reported by 
Inglis \textit{et. al.}~\cite{inglis_fluidic_2020}, wherein $N_p=N_w=8$, $g_w=g_a=7~\mu m$, and $\lambda_w=\lambda_a=20~\mu m$.
We performed fluid flow simulations with various mesh sizes.
We conducted particle tracking simulations and obtained variations of critical diameter throughout channel for each case.
We considered a few metrics when examining the results obtained from different grids and observed that mesh convergence can be achieved with a mesh size of $\delta\approx0.05 g_w$.
In particular, the local critical diameter in the center of channel varied about $16.76\%$ when doubling the mesh size, while the corresponding value was about $0.96\%$ when halving the mesh size.
The corresponding values for variations of mean absolute error (MAE) related to local critical diameter matrices were about $378.8$ nm, and $39.9$ nm, respectively.
A typical mesh in this work consisted of approximately 900,000 and 1,000,000 cells.
We used a computer with 
16.0 GB RAM and a 
13th Gen Intel\textsuperscript\textregistered~Core\textsuperscript{TM} i9-13980HX Processor.
A typical fluid flow simulation took approximately 20 to 45 minutes to complete, depending on the case complexity and system workload.

\subsection{Pressure balance}
\label{sec_pressure_balance}

The pressure balance model developed by Inglis \textit{et. al.}~\cite{inglis_fluidic_2020} 
aims at configuring the interface unit cell next to accumulation sidewall.
The subject unit cell 
is shown 
in the boxed region next to accumulation sidewall
in Fig.~\ref{fig_schem}, 
and
borders the accumulation sidewall,
upstreammost 
row of a given full DLD unit (row \#$0$ with a 0-based index notation), and
downstreammost 
row of neighboring full DLD unit upstream of the full DLD unit (row \#$N_p-1$ with a 0-based index notation).
The model can be described as shown in Eq.~\ref{eq_perss_balance_inglis}:
\begin{equation}
	R_w^\text{Int.}+R_a^\text{Int.}=R_a(1+\varepsilon),
	\label{eq_perss_balance_inglis}
\end{equation}
wherein $R_w^\text{Int.}$ and $R_a^\text{Int.}$ denote the lateral and axial resistances of interface unit cell, 
$R_w$ and $R_a$ refer to the lateral and axial hydraulic resistances of a unit cell in the bulk of array, 
and
$\varepsilon=1/N_p$ is the row shift fraction.
To the best of our knowledge, 
no guidelines are currently available for determining 
the optimal values for the paired resistances of
$R_w^\text{Int.}$ and 
$R_a^\text{Int.}$.

One of the goals of this work was to examine a wide spectrum of possibilities of the paired resistances of
$R_w^\text{Int.}$ and 
$R_a^\text{Int.}$ 
for DLD systems with various values of periodicity.
In order to accomplish that, we introduce a new parameter $\phi$ in our implementation defined as
\begin{equation}
	\phi\equiv\frac{R_w^\text{Int.}}{R_a^\text{Int.}}.
	\label{eq_phi}
\end{equation}
The pressure balance model, Eq.~\ref{eq_perss_balance_inglis}, in conjunction with Eq.~\ref{eq_phi}, can be rewritten as
\begin{equation}
	\frac{R_a^\text{Int.}}{R_a}=\frac{1+\varepsilon}{1+\phi}.
	\label{eq_press_balance_implementation}
\end{equation}
For a given $\phi$, the axial hydraulic resistance of interface unit cell next to accumulation sidewall $R_a^\text{Int.}$ can be found from Eq.~\ref{eq_press_balance_implementation}, which then gives the corresponding lateral gap of unit cell $g_w^\text{Int.}$.
The lateral hydraulic resistance of interface unit cell next to accumulation sidewall $R_w^\text{Int.}$ can be found from Eq.~\ref{eq_phi} afterwards, which can be used to determine the corresponding axial gap of the unit cell $g_a^\text{Int.}$.

\begin{figure*}[!tb]
	\centering
	\includegraphics[width=\textwidth]{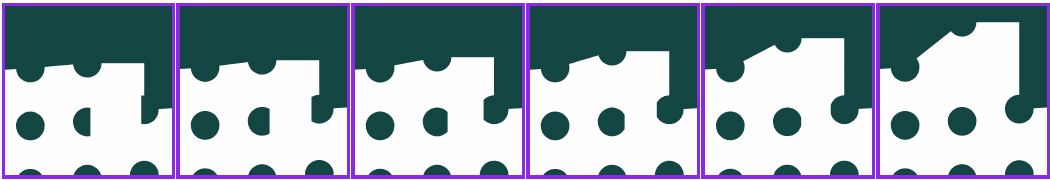}
	\caption{Interface unit cell similar to that shown in the boxed region next to accumulation sidewall in Fig.~\ref{fig_schem}
		for
		a DLD system with $N_p=14$ and $g_w=g_a=12.7~\mu m$ designed with various values of pressure balance parameter, from left to right, $\phi=0.25,~0.5,~1.0,~2.0,~5.0,~\text{and}~10.0$.
	}
	\label{fig_intro_various_phi}
\end{figure*}
We have shown the geometrical configuration of interface unit cell of
a DLD system with $N_p=14$ and $g_w=g_a=12.7~\mu m$ designed with various values of pressure balance parameter in Fig.~\ref{fig_intro_various_phi}.
It can be discerned that 
the axial gap increases and lateral gap decreases as we reduce $\phi$.
The axial gap widening is accomplished by cutting adjacent pillars as explained in Inglis \textit{et. al.}~\cite{inglis_fluidic_2020}.
The level of cutting pillars decreases as $\phi$ increases, necessitating the lateral gap to increase sufficiently to hold the pressure balance model.

This pressure balance scheme has been implemented in the DDA tool so that 
different sidewall configurations can be designed for various $\phi$ values provided by user, through determining appropriate axial and lateral resistances and by applying the required gap adjustments automatically.
Here, we use the data-driven model developed by Inglis \textit{et. al.}~\cite{inglis_fluidic_2020} to calculate the required resistances.
One caveat, however, is that the model is mainly developed for fluid flow around cylinder, which may not be valid for evaluation of lateral resistance with pillars being partially cut.
The $\sim$90-degree turn of fluid flow within the interface unit cell and its corresponding energy losses 
are other factors that may not be captured by this model accurately.
Thus, while the pressure balance mechanism serves as a guideline, its limitations necessitate a thorough investigation of various feasible configurations to identify an optimal solution.
As such an optimal solution may depend on the configurations of bulk of the array, we will study DLD systems with various values of periodicity in Section~\ref{sec_res_periodicity}.

%% file: results.tex
\section{Results and Discussion}
\label{sec_res}

\subsection{Addressing unit cells in DLD array} 
\begin{figure*}[!tb]
	\centering
	\includegraphics[width=\textwidth]{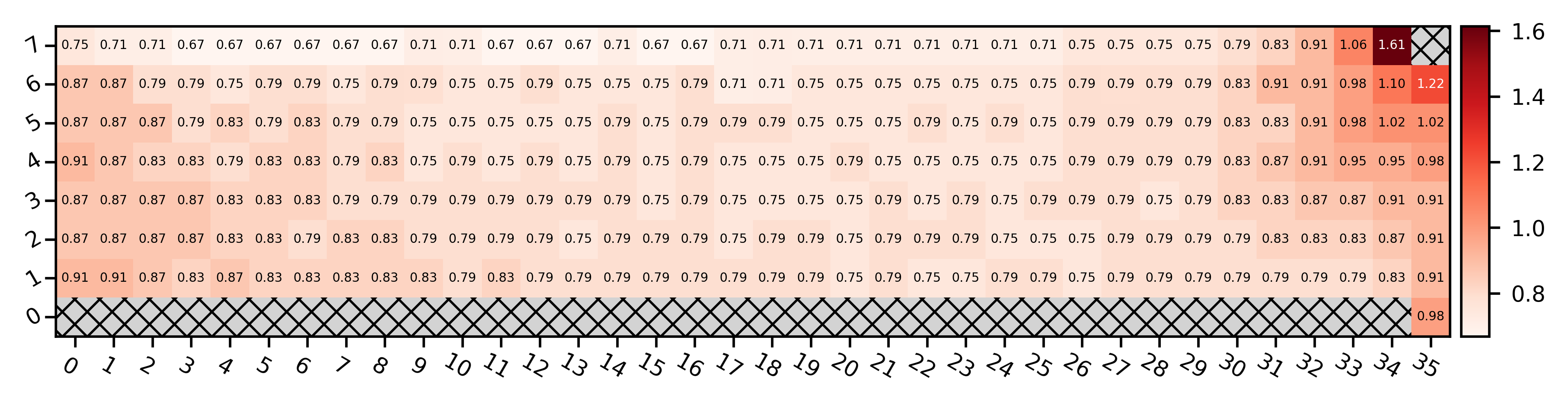}
	\includegraphics[width=\textwidth]{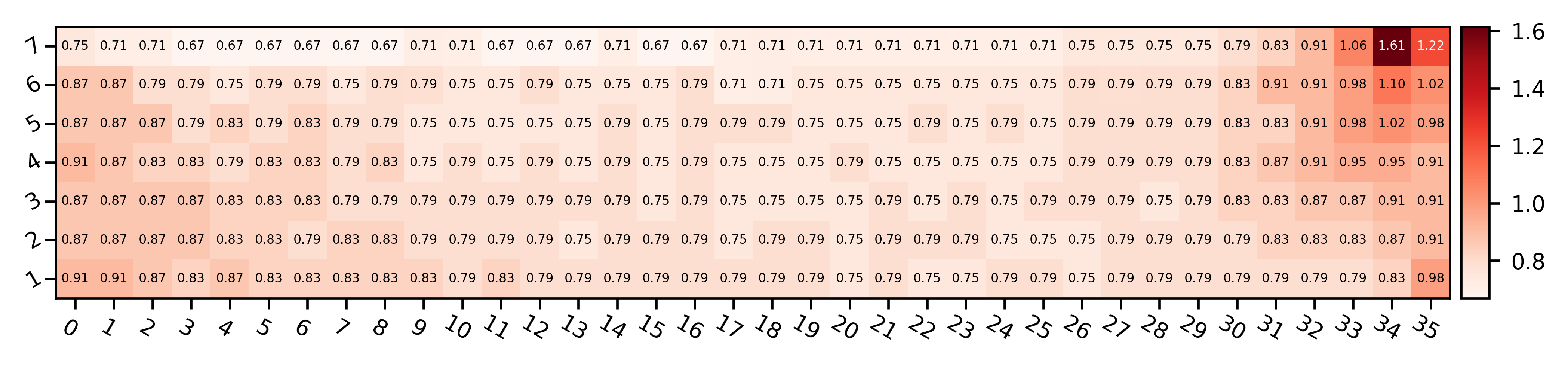}
	\caption{Variations of estimated dimensionless critical diameter, $d_c^\ast\approx 2\beta/d_c^\text{Poiseuille}$, throughout a full DLD unit for a system with $N_p=36$, $N_w=8$, $g_w=g_a=7~\mu m$, and $\lambda_w=\lambda_a=14~\mu m$, wherein the boundary profiles are designed by using the \codex1{3d} model with a pressure balance parameter of $\phi=1$, presented in two different styles: full (top) and compact (bottom).
		The unit cells wherein the concepts of first stream width and local critical diameter are not applicable are hatched with lines in the full style, while they are omitted in the compact style in conjunction with an upward shift in unit cells residing on Row \#$N_p-1$.
		The theoretical critical diameter is estimated to be $d_c^\text{Poiseuille}\approx1.39~\mu m$~\cite{inglis_critical_2006}.
	}
	\label{fig_res_full_vs_compact}
\end{figure*}

A standardized approach to the presentation of results, 
\textit{e.g.}, variations of FSFF, $\beta$, and derivative variables, 
is notably absent in the literature.
In addition, it seems to be common to use only Row indices when reporting these variables, which can cause confusion about the exact location of unit cells within the pillars array.
These issues can make it challenging for researchers to reproduce the pertinent works or to use a reported dataset in their own work.
Here, prior to presenting and discussing our main results, we need to provide some information on our notation and data presentation approach.

By way of illustration, consider a full DLD unit of periodicity of $N_p=36$ with 8 fluidic lanes, \textit{i.e.}, $N_w=8$.
The obtained results for dimensionless critical diameter variations throughout a full DLD unit are shown in Fig.~\ref{fig_res_full_vs_compact} 
for the case that the boundary profiles are designed by using the \codex1{3d} model with a pressure balance parameter of $\phi=1$.
We present our results in two fashions: 1. full, and 2. compact. 
In \textit{full} style, the unit cells wherein the concepts of first stream width and local critical diameter are not applicable are hatched with lines.
We omit these cells for the \textit{compact} style while applying an upward shift in unit cells residing on Row \#$N_p-1$, which reduces the matrix size from $N_p\times N_w$ to $N_p\times (N_w-1)$.
While the \textit{compact} style benefits from its conciseness, the \textit{full} format is likely more intuitive and comprehensible.
Unless otherwise stated, we use the \textit{full} format in this work.

\subsection{Validation}
\label{sec_res_validation}

\begin{figure}[!tb]
	\centering
	\includegraphics[width=0.9\textwidth]{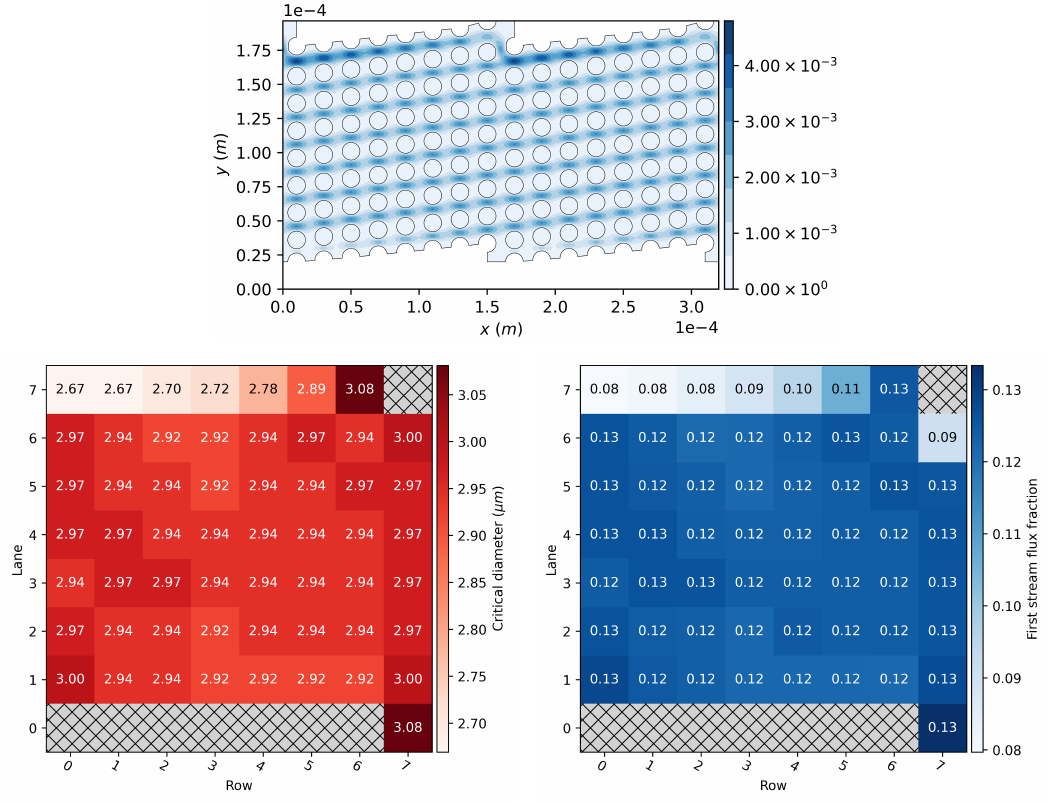}
	\caption{Velocity field (m/s) obtained from numerical simulation (top) 
		together with 
		variations of
		local critical diameter (bottom left)
		and 
		first stream flux fraction (bottom right)
		across a full DLD unit for a system similar to that reported by Inglis \textit{et. al.}~\cite{inglis_fluidic_2020}: $N_p=N_w=8$, $g_w=g_a=7~\mu m$, and $\lambda_w=\lambda_a=20~\mu m$.
		The ideal critical diameter is estimated to be $d_c^\text{Poiseuille}\approx 3.09~\mu m$~\cite{inglis_critical_2006}.
		The pressure balance parameter is set to $\phi=1.79$, which leads to the lateral and axial gaps of pillars in unit cell $(7,7)$ being equal to $10~\mu m$ and $8~\mu m$, respectively, \textit{i.e.},~$3~\mu m$ lateral gap widening together with $1~\mu m$ axial gap widening achieved through cutting the adjacent pillars similar to what shown in the boxed region next to the accumulation sidewall in Fig.~\ref{fig_schem}.
		512 tracers were used within each gap for particle tracking simulations, which gives an uncertainty of ${\sim}\pm 13.7$ nm for the calculated critical diameters.
 	}
	\label{fig_res_benchmark_Inglis}
\end{figure}

Here, in order to validate our code implementation, we consider a DLD system similar to that reported by 
Inglis \textit{et. al.}~\cite{inglis_fluidic_2020}.
The bulk configurations include $N_p=N_w=8$, $g_w=g_a=7~\mu m$, and $\lambda_w=\lambda_a=20~\mu m$. 
We set the pressure balance parameter to $\phi=1.79$.
This gives the lateral and axial gaps of pillars of the unit cell $(7,7)$ at the interface of full DLD units, 
similar to what shown in the boxed region next to accumulation sidewall in Fig.~\ref{fig_schem},
to be $g_w^\text{Int.}=10~\mu m$ and $g_a^\text{Int.}=8~\mu m$, respectively.
That is, 
knowing that the gap is $g_w=7~\mu m$ in the bulk of array,
a~$3~\mu m$ widening is applied to lateral gap, while a $1~\mu m$ axial gap widening is accomplished through cutting the corresponding adjacent pillars.
It is worth mentioning that in the work of
Inglis \textit{et. al.}~\cite{inglis_fluidic_2020}, 
the axial gap was $9~\mu m$.
We think that might be due to what seems to be a mistake in their implemented code accessible from the ESI of the original work, wherein 
the point that 
cutting a pair of pillars by a given amount, \codex1{R\_cut}, results in a gap widening twice as large as \codex1{R\_cut}
seems to have been overlooked.

The obtained results are shown in Fig.~\ref{fig_res_benchmark_Inglis}.
As we have used 512 tracers within each gap for these particle tracking simulations, the uncertainty of the calculated critical diameters is ${\sim}\pm 13.7$ nm.
It can be discerned that the largest value of critical diameter is ${\sim} 3.08\pm 0.01~\mu m$, which occurs next to the sidewalls.
The local critical diameter at the center of channel is ${\sim}2.94\pm 0.01~\mu m$.
We introduce $\Delta\psi\equiv {(d_c-d_c^\text{mid-channel})}/{d_c^\text{mid-channel}}$ as a nondimensional metric to measure how large a local critical diameter ($d_c$) is compared to that in the center of channel ($d_c^\text{mid-channel}$).
The maximum of this metric over Lanes \#0 (depletion sidewall) and \#7 (accumulation sidewall) is ${\sim}4.6\%$.
The maximum value for Lanes \#3 and 4 (central lanes) is ${\sim}0.9\%$.
The boundary treatment can, therefore, be considered effective.
In fact the results seem to be even more promising than those reported by 
Inglis \textit{et. al.}~\cite{inglis_fluidic_2020}, as the maximum critical diameter in the original work was reported to be ${\sim}3.98~\mu m$.
The maximum critical diameter in our simulation, ${\sim} 3.08\pm 0.01~\mu m$, is ${\sim}23\%$ smaller.
The discrepancy may be attributed to the fact that in the original work, a three-dimensional system was studied, while we have conducted a 2D modeling.
Although the pillars height was relatively large in the original work, $h^\ast=2$, the corresponding boundary gap profile is still slightly different from that of ``infinitely-deep'' channels that can be represented by systems with $h^\ast\ge2.5$ according to Fig.~\ref{fig_res_boundary_gaps}.
The other factor that may explain the observed discrepancy is the difference in axial gaps of the interface unit cells employed in the two works as discussed above.
In fact, as we will show in Section~\ref{sec_res_periodicity}, the geometrical configurations of interface unit cell next to accumulation sidewall can affect the results noticeably.
Lastly, the obtained results also show a relatively uniform FSFF across the channel with slight variations around the ideal value ($1/Np=0.125$).

\subsection{Effects of pressure balance parameter and periodicity}
\label{sec_res_periodicity}

\begin{figure}
	\centering
	\includegraphics[width=0.965\textwidth]{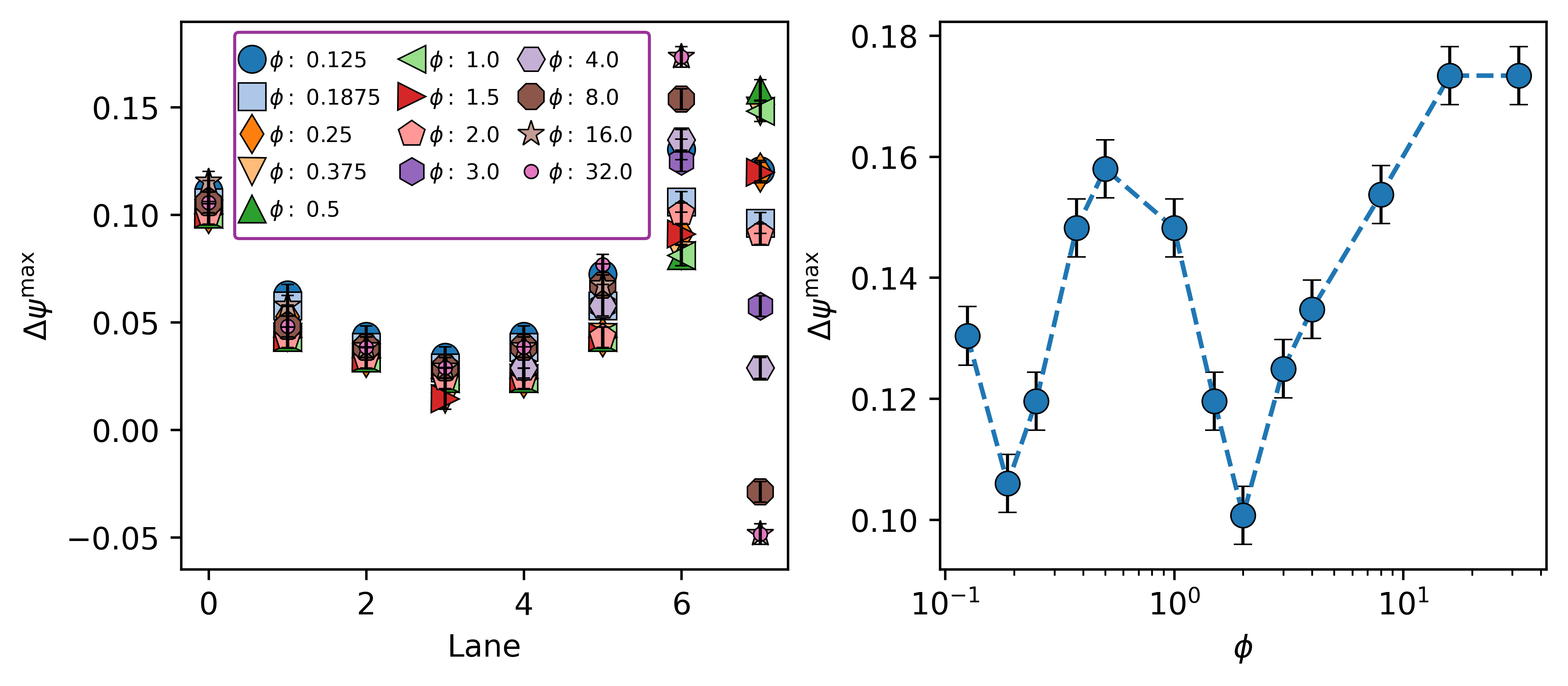}
	\includegraphics[width=0.965\textwidth]{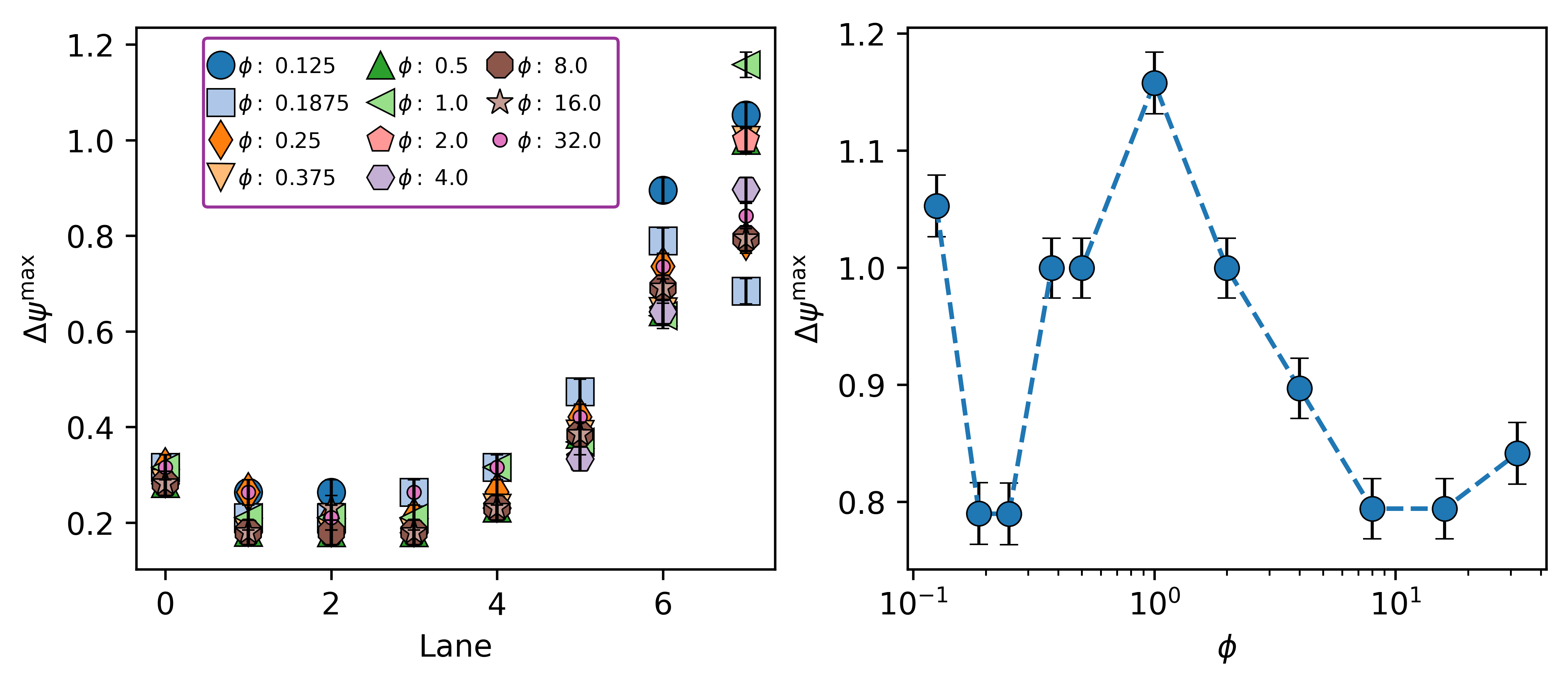}
	\includegraphics[width=0.965\textwidth]{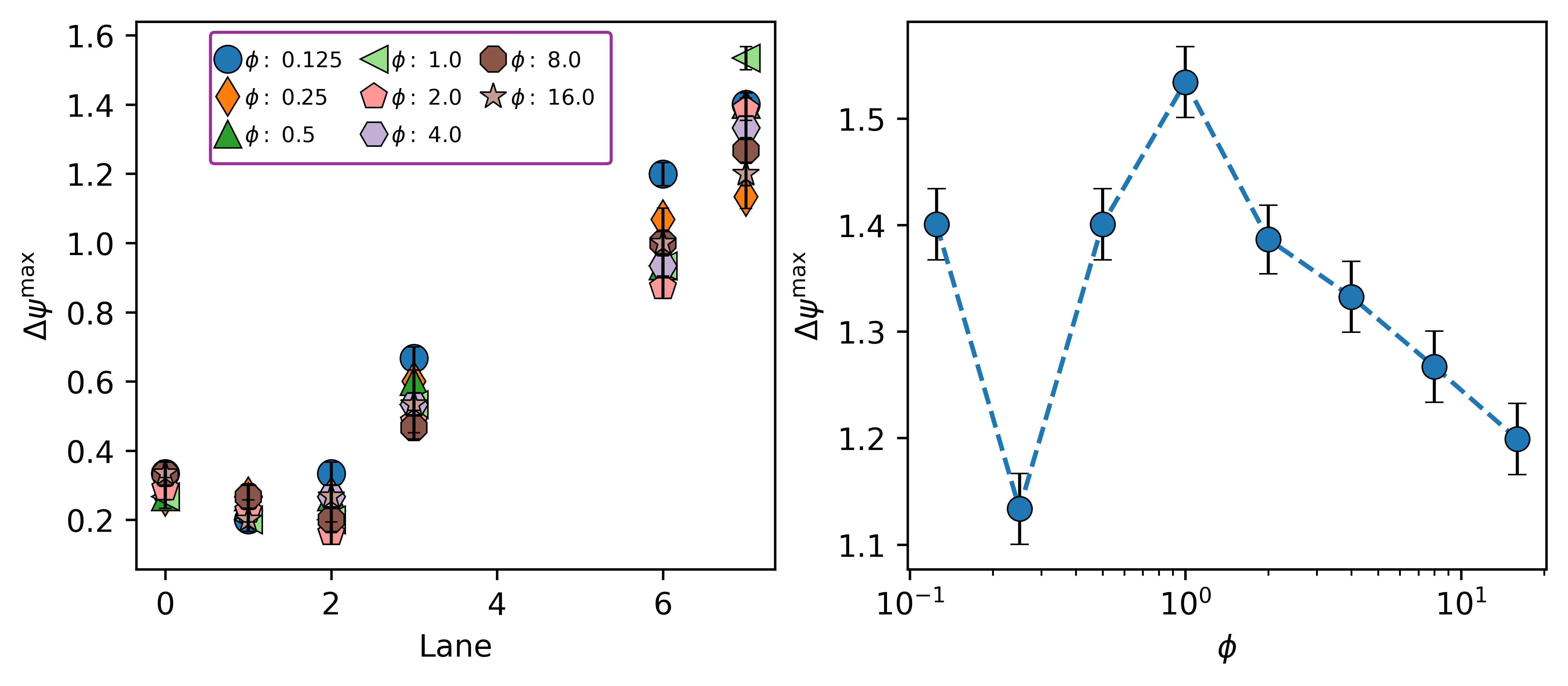}
	\caption{Maximum nondimensional deviation of critical diameter in each lane (left) 
	and that of the entire DLD unit (right) for 
	a system with
	$g_w=g_a=7~\mu m$, $\lambda_w=\lambda_a=14~\mu m$
	and $N_p=8$ (top), $N_p=36$ (middle), and  $N_p=50$ (bottom)
	obtained by applying the gap profile from the model of Inglis \textit{et. al.}~\cite{inglis_fluidic_2020} in conjunction with various values of pressure balance parameter $\phi$.
	The error bars show the uncertainty due to the fact that a finite number of tracers has been used for particle tracking simulations to obtain first stream width.
	Note that particle tracking has not been done for Lanes \#4 and 5 of systems with $N_p=50$ to manage computational costs.
	}
	\label{fig_res_various_np_phi}
\end{figure}

In this section, we explore a relatively wide spectrum of pressure balance parameter $\phi$ and DLD periodicity $N_p$ to investigate the fluid flow characteristics across channel.
We consider systems with $N_w=8$, $g_w=g_a=7~\mu m$, $\lambda_w=\lambda_a=14~\mu m$
and three periodicity levels of $N_p=8,~36,~\text{and}~50$, for each set, several pressure balance parameters are studied over the range of 0.125 to 32.
We consider 512 tracers per gap for particle tracking simulations in systems with $N_p=8$.
For longer systems ($N_p>8$), we use 256 tracers per gap to manage computational costs.

The results are shown in Fig.~\ref{fig_res_various_np_phi}.
There are several findings that are listed in the following.

\begin{itemize}
	\item Lanes closer to sidewalls have larger maximum-over-lane critical diameter. This effect is typically more noticeable for accumulations sidewall.
	\item 
	Based on the pressure balance model described in Section~\ref{sec_pressure_balance}, one might assume that boundary profiles generated by varying $\phi$ would perform identically
	as the axial ($R_a^\text{Int.}$) and lateral ($R_w^\text{Int.}$) hydraulic resistances are determined in such a way that Eq.~\ref{eq_perss_balance_inglis} is satisfied.
	However, as shown in Fig.~\ref{fig_res_various_np_phi}, the parameter $\phi$ significantly influences the characteristics.
	This is primarily because the model relies on an empirical formula developed by Inglis \textit{et. al.}~\cite{inglis_fluidic_2020} to calculate the required resistances.
	The empirical formula is derived for fluid flow around a cylinder in a periodic structure of infinite extent.
	In practice, the empirical formula may not accurately determine the axial and lateral hydraulic resistances of interface unit cell for several reasons. 
	First, the assumption of a periodic system becomes invalid for the interface unit cell as it borders the accumulation sidewall.
	Second, the geometrical configuration of interface unit cell does not perfectly align with that used to derive the empirical formula: a) the cut pillars can alter the fluid flow pattern compared to that around a full circular cylinder, and b) the profile of sidewall close to interface unit cell is not parallel with the channel axis.
	Third, the fluid flow shows a relatively complex pattern within the interface unit cell and has a $\sim$90-degree turn, the resistances of which may not be accurately estimated by the empirical formula.
	We should also clarify that, given the aforementioned complexities, 
	estimating
	the net resistance of the interface unit cell as two resistances connected in series ($R_a^\text{Int.}+R_w^\text{Int.}$) may contain inherent errors in the first place, but we still use it in the absence of a more precise model.
	\item It can be discerned that there are generally two optimal ranges for this parameter; one is smaller than 1, and the other is larger than 1.
	The former is related to the case that a significant portion of adjacent pillars in interface unit cell needs to be removed and lateral gap is not widened significantly, while the latter is associated with the opposite scenario; minimally cutting the adjacent pillars of interface unit cell and more noticeably widening the lateral gap.
	\item The pressure balance parameter of $\phi\approx 0.1875$ points to one of the optimal parameter regions regardless of periodicity.
	The second optimal solution is larger than 1, but is periodicity-dependent, \textit{e.g.}, 
	$\phi\approx 2.0$ for the case of $N_p=8$, and 
	$\phi\approx 10.0$ for the case of $N_p=36$.
	\item Manufacturing considerations should be taken into account when choosing a $\phi$ value.
	It may be challenging to manufacture high aspect ratio pillars with small values of $\phi$.
	On the other hand, the lateral gap between pillars of interface unit cells can be noticeably altered for large values of $\phi$, which can cause local depth non-uniformity during etch processes like reactive ion etch (RIE).
\end{itemize}

\subsection{Various models for boundary gap profiles}
\label{sec_res_gap_profile}

\begin{figure}
	\centering
	\includegraphics[width=0.8\textwidth]{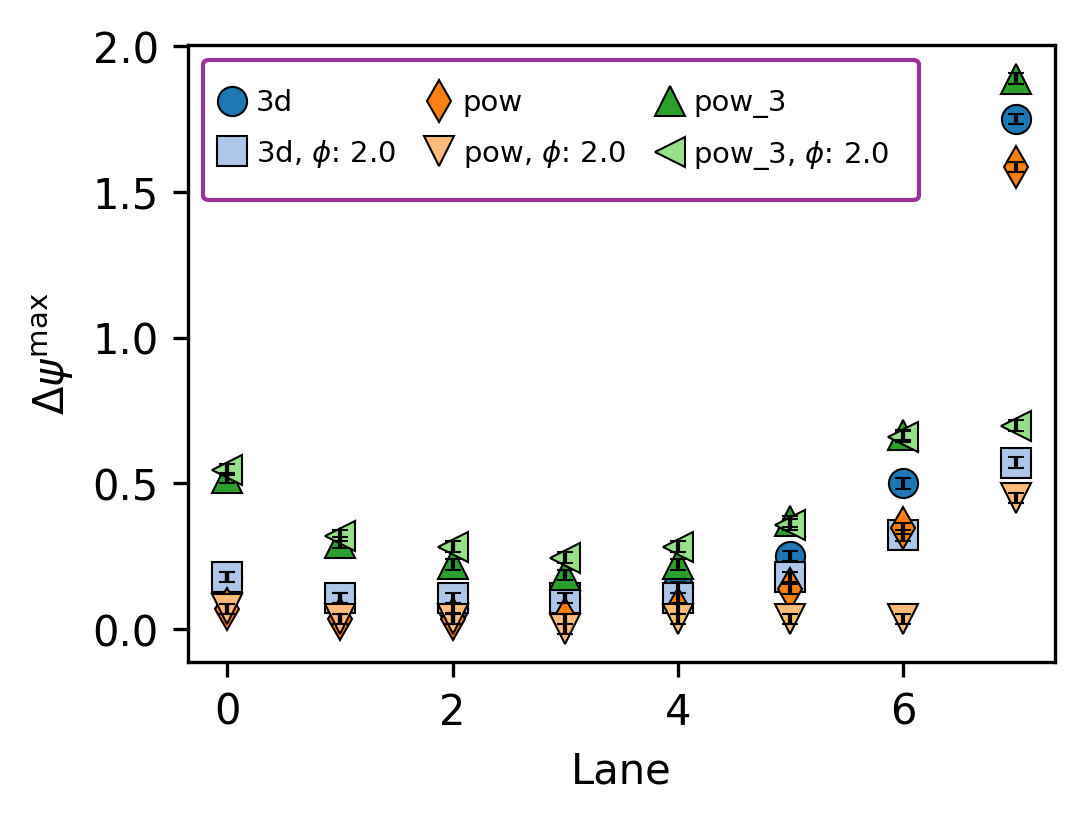}
	\caption{Maximum nondimensional deviation of critical diameter in each lane of a DLD system with
		$N_p=22$, $N_w=8$, $g_w=g_a=7~\mu m$, and $\lambda_w=\lambda_a=14~\mu m$ obtained by applying different boundary treatment approaches:
		\codex1{pow\_3}~\cite{feng_maximizing_2017},
		\codex1{pow}~\cite{ebadi_efficient_2019}, and
		\codex1{3d}~\cite{inglis_fluidic_2020},
		with ($\phi=2$) and without pressure balance adjustment.
		The error bars show the uncertainty due to the fact that a finite number of tracers has been used for particle tracking simulations to obtain first stream width.
	}
	\label{fig_res_compare_methods}
\end{figure}

\begin{table*}[!tb]
\centering
\caption{Maximum nondimensional deviation of critical diameter $\Delta\psi^\text{max}$ for a DLD system with
	$N_p=22$, $N_w=8$, $g_w=g_a=7~\mu m$, and $\lambda_w=\lambda_a=14~\mu m$ obtained by applying different boundary treatment approaches:
	\codex1{pow\_3}~\cite{feng_maximizing_2017},
	\codex1{pow}~\cite{ebadi_efficient_2019}, and
	\codex1{3d}~\cite{inglis_fluidic_2020},
	with ($\phi=2$) and without pressure balance adjustment.
	The uncertainty of values due to using a finite number of tracers in particle tracking simulations is ${\sim}\pm 0.02$.
	}
\begin{tabular*}{\textwidth}{@{\extracolsep{\fill}}||l c c c||} 
	\hline
	Model & \codex1{3d} & \codex1{pow} & \codex1{pow\_3} \\ [0.5ex] 
	\hline\hline
	Without pressure balance & $1.75$ & $1.59$ & $1.89$ \\ 
	With pressure balance & $0.57$  & $0.45$ & $0.70$ \\  [1ex] 
	\hline
\end{tabular*}
\label{tab_max_metric}
\end{table*}

\begin{figure}[p]
	\centering
	\includegraphics[width=\textwidth]{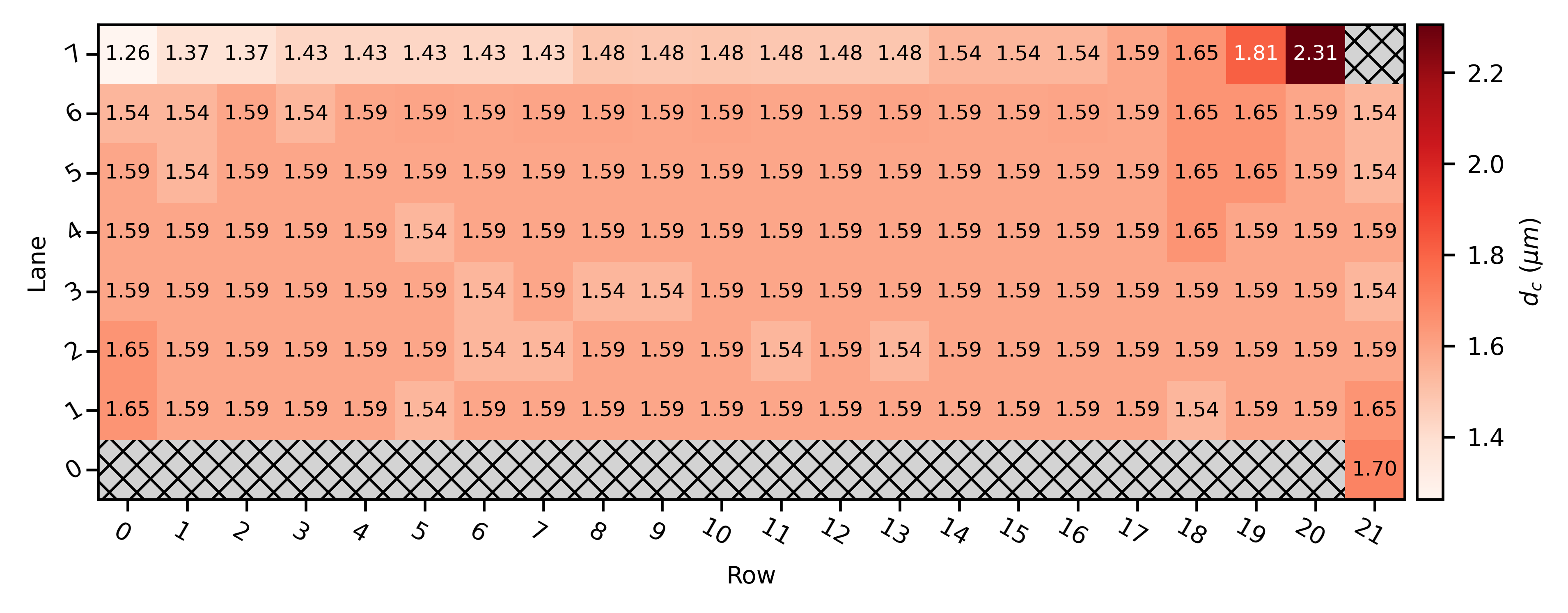}
	\includegraphics[width=\textwidth]{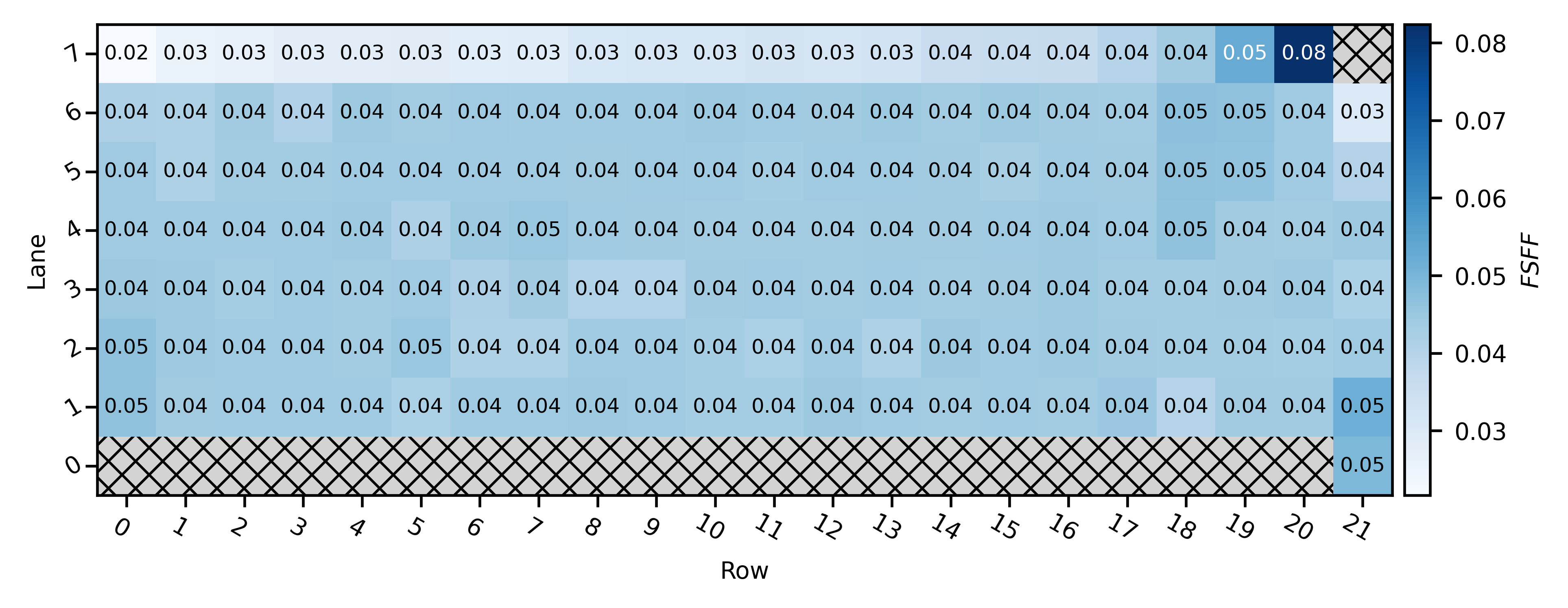}
	\includegraphics[width=\textwidth]{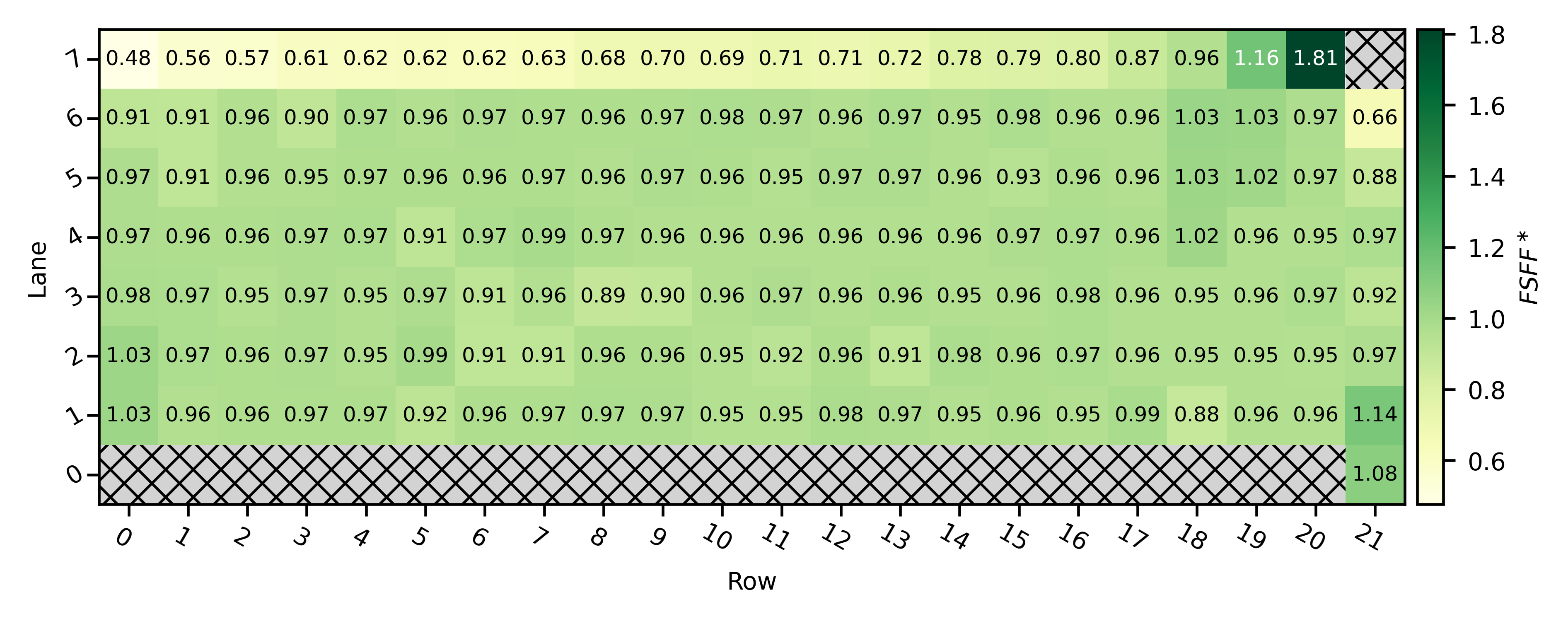}
	\caption{Variations of estimated critical diameter $d_c\approx 2\beta$ (top) together with dimensional (middle) and nondimensional (bottom) FSFF
	across channel for a DLD system with
	$N_p=22$, $N_w=8$, $g_w=g_a=7~\mu m$, and $\lambda_w=\lambda_a=14~\mu m$ obtained by applying the boundary treatment approach reported by Ebadi \textit{et. al.}~\cite{ebadi_efficient_2019} integrated with pressure balance mechanism~\cite{inglis_fluidic_2020} with a resistance ratio of $\phi=2$.
	The ideal critical diameter is estimated to be $d_c^\text{Poiseuille}\approx 1.8~\mu m$~\cite{inglis_critical_2006}.
	}
	\label{fig_res_pow}
\end{figure}

In the DDA tool, we have implemented the pressure balance scheme so that it can be integrated into any boundary treatment approach flexibly.
Here, we explore the performance of models from the works of
Feng \textit{et. al.}~\cite{feng_maximizing_2017},
Ebadi \textit{et. al.}~\cite{ebadi_efficient_2019}, and
Inglis \textit{et. al.}~\cite{inglis_fluidic_2020}
for designing sidewalls with and without pressure balance mechanism.
We consider systems with 
$N_p=22$, $N_w=8$, $g_w=g_a=7~\mu m$, and $\lambda_w=\lambda_a=14~\mu m$.
The ideal critical diameter is estimated to be $d_c^\text{Poiseuille}\approx 1.8~\mu m$~\cite{inglis_critical_2006}.
It is worth mentioning that herein, we do not attempt to find the optimal $\phi$ for each model.
Instead, we use the same $\phi=2$ when applying the pressure balance scheme.

The obtained results are shown in Fig.~\ref{fig_res_compare_methods}.
It can be perceived that all models can significantly benefit from the pressure balance mechanism.
It can also be discerned that applying pressure balance mechanism has much more noticeable effects on the half of channel closer to the accumulation sidewall.
In particular, the first few rows in proximity of accumulation sidewall experience the most significant impact.
For all the cases studied in this section, the maximum nondimensional deviation of critical diameter $\Delta\psi^\text{max}$ occurs in the lane adjacent to  the accumulation sidewall.
The corresponding values are reported in Table~\ref{tab_max_metric}.
The nonuniformity reduces from 
${\sim}175\%$ to ${\sim}57\%$ for \codex1{3d} model~\cite{inglis_fluidic_2020},
${\sim}159\%$ to ${\sim}45\%$ for \codex1{pow} model~\cite{ebadi_efficient_2019}, and
${\sim}189\%$ to ${\sim}70\%$ for \codex1{pow\_3} model~\cite{feng_maximizing_2017}.

\begin{figure}[!bt]
	\centering
	\includegraphics[width=\textwidth]{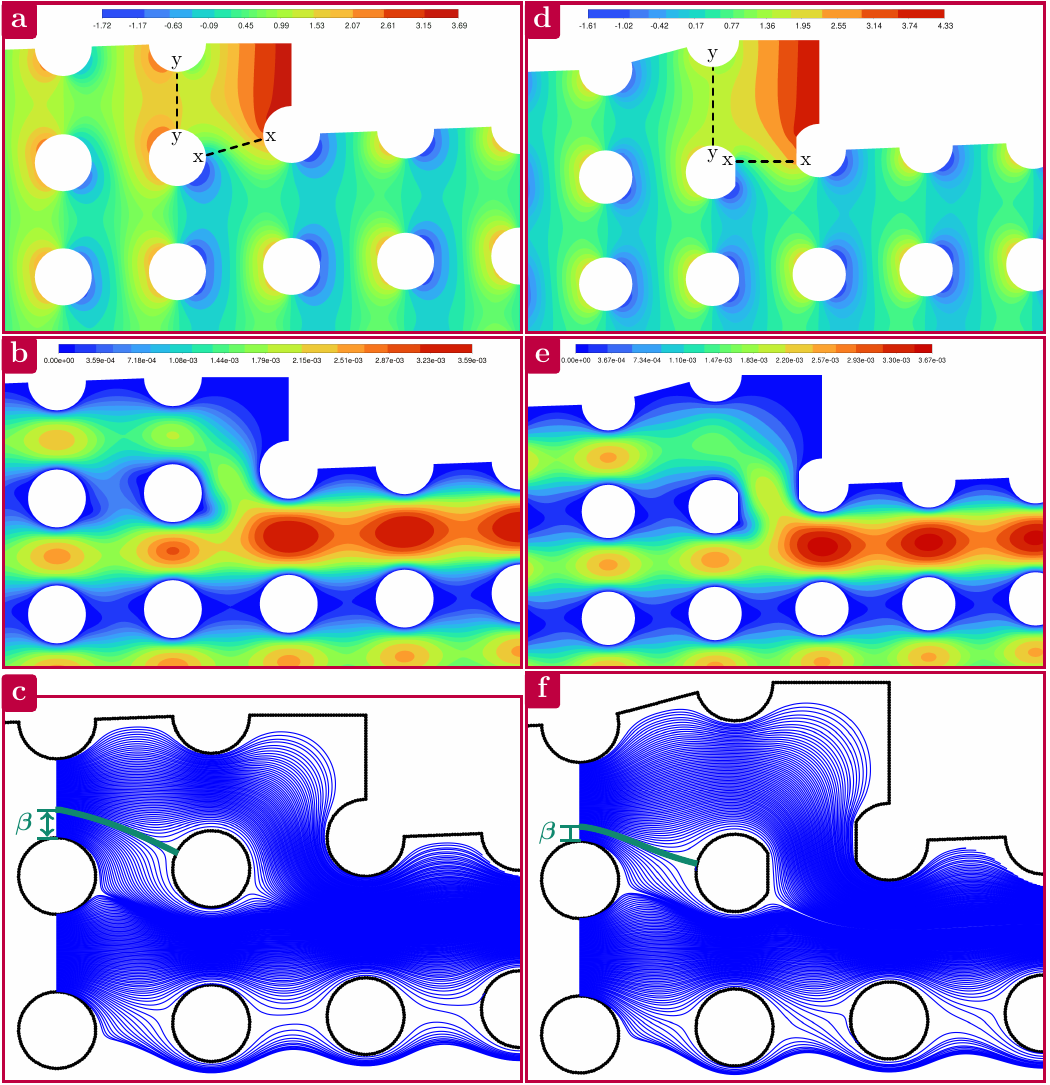}
	\caption{Variations of pressure in Pa unit (a, d), velocity magnitude in m/s unit (b, e), together with streamline patterns (c, f) inside and around interface unit cell close to accumulation sidewall similar to what shown in the boxed region in Fig.~\ref{fig_schem} 
	for a DLD system with
	$N_p=22$, $N_w=8$, $g_w=g_a=7~\mu m$, and $\lambda_w=\lambda_a=14~\mu m$
	wherein the boundary gap profile is determined by using the \codex1{pow} model without (a\textendash c) and with (d\textendash f) pressure balance.
	The pressure balance parameter is $\phi=2$ in (d\textendash f).
	The first stream width shown in c and f is about $\beta=2.06~\mu m$ and $\beta=1.155~\mu m$, respectively.
	}
	\label{fig_res_interface}
\end{figure}

\begin{figure}[!bt]
	\centering
	\includegraphics[width=\textwidth]{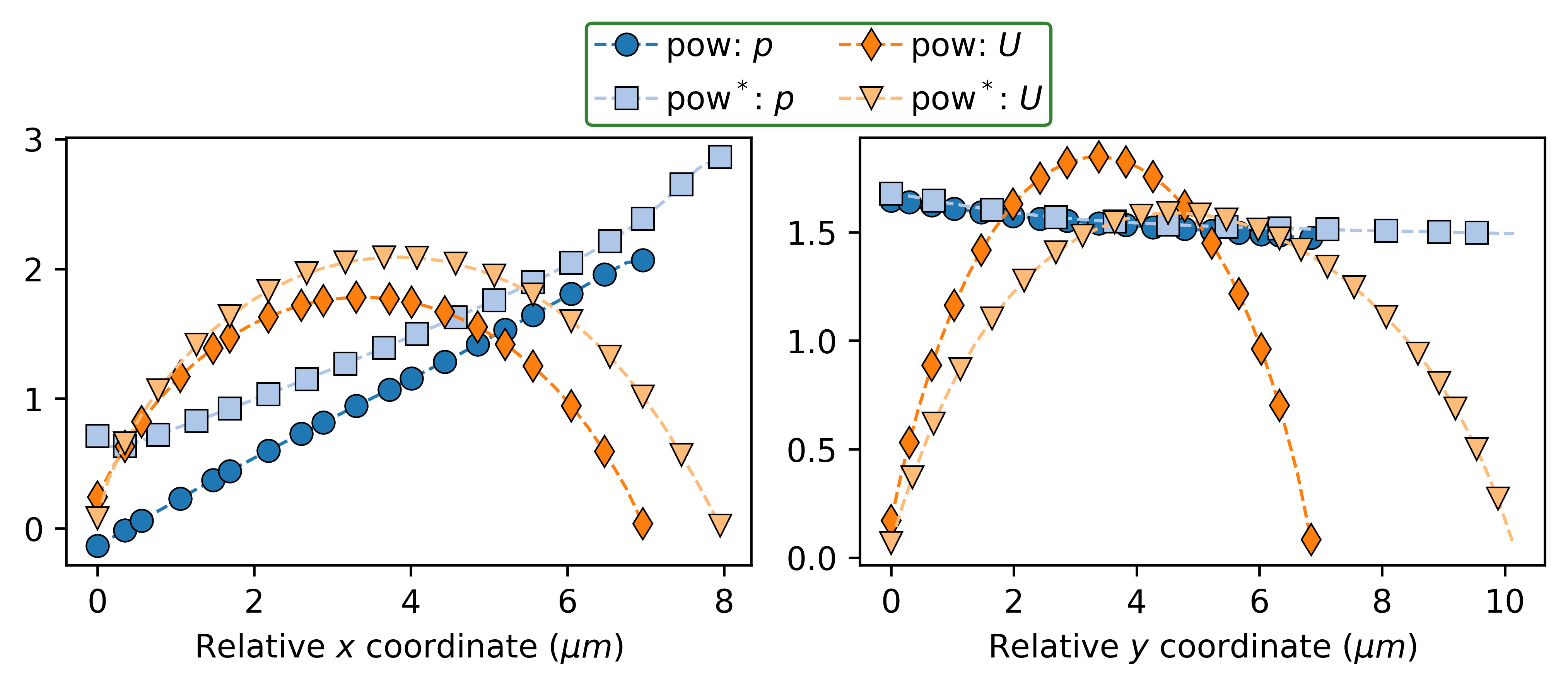}
	\caption{Variations of pressure $p$ (Pa) and velocity magnitude $U$ (mm/s)
		along the x-x (left) and y-y (right) sections shown in Fig.~\ref{fig_res_interface}
		for the case that boundary gap profile is determined by using the \codex1{pow} model without and with (\codex1{pow$^\ast$}) applying the pressure balance using $\phi=2$.
	}
	\label{fig_res_interface_xy}
\end{figure}

Another observation is that the \codex1{pow\_3} model shows the highest nonuniformity.
Even the depletion sidewall seems to have not been treated effectively as it shows a maximum nonuniformity of 
${\sim}51.8\%$ and 
${\sim}54.7\%$ in Lane \#0  without and with pressure balance scheme, respectively.
The maximum nonuniformity of 
\codex1{3d} model in Lane \#0 is obtained to be ${\sim}17.9\%$ regardless of pressure balance.
These values decrease to ${\sim}6.9\%$ for 
\codex1{pow} model regardless of pressure balance.

Overall, we found the integration of \codex1{pow} model~\cite{ebadi_efficient_2019} and pressure balance model~\cite{inglis_fluidic_2020} to result in the least nonuniformity.
The variations of estimated critical diameter ($2\beta$), together with FSFF and its equivalent form nondimensionalized by its ideal counterpart, \textit{i.e.}, $1/N_p\approx 0.045$
are shown in Fig.~\ref{fig_res_pow}.

We have illustrated the pressure and velocity contours as well as streamline patterns within and close to the interface unit cell adjacent to accumulation sidewall in Fig.~\ref{fig_res_interface} for the \codex1{pow} model with and without the pressure balance mechanism.
The results show that
widening the axial and lateral gaps of the interface unit cell through
employing the pressure balance scheme with $\phi=2$ causes a pressure rise of up to $\sim 17\%$ close to the boundary between the interface unit cell and accumulation sidewall. 
The variations of pressure and velocity magnitude along the axial and lateral gaps are shown in Fig.~\ref{fig_res_interface_xy}.
In particular, by increasing the gaps of interface unit cell and reducing its net hydraulic resistance through applying the pressure balance with $\phi=2$, the volumetric flow rate into the cell increases by $\sim 30\%$,
which directly affects the fluid flow characteristics of neighboring cells.
For example, 
with regard to the upstream neighboring cell,
it causes a reduction of fluid flux into the lower fluidic lane (zigzag mode) so it can supply $\sim 30\%$ more fluid into the interface unit cell,
which reduces the corresponding first stream width approximately from $\beta=2.06~\mu m$ to $\beta=1.15~\mu m$ as shown in Fig.~\ref{fig_res_interface} c and f.

%% file: usage.tex

\section{Applying the findings: Practical guidance}
\label{sec_practical_guide}

The obtained results in Section~\ref{sec_res_gap_profile} demonstrated that 
regardless of the model used for determining the boundary
gap profiles, applying the pressure balance scheme can reduce the critical diameter variations and improve the separation performance.
These findings provide profound insight into how to develop effective methods for designing the boundary profiles of DLD system, offering a clear pathway for future research.
In particular, any model should consist of at least two components:
1. an appropriate boundary gap profile for Rows \#$0$ to $N_p-2$, and
2. a suitable pressure balance mechanism for unit cell at Row \#$N_p-1$.

Researchers can use these findings in their works in a few different ways as explained in the following.
\begin{itemize}
	\item After incorporating the required modifications pointed out in Section~\ref{sec_res_validation}, one can use the implementation provided in the ESI of the work reported by Inglis \textit{et. al.}~\cite{inglis_fluidic_2020}, to apply the pressure balance in conjunction with any currently available model, or novel methods that they may develop independently, for determining the boundary gap profiles.
	
	\item One can also follow the implementation process described in Section~\ref{sec_pressure_balance} to develop their own pressure balance module.
	
	\item It is also feasible to use our implemented code which is accessible from the {\mnflow} package~\cite{mehboudi_mnflow_2024}.
	
	\item The previous options can provide flexibility and are suitable for developing new methodologies.
	For currently available models, however, researchers can alternatively use the DDA tool from the {\mnflow} package~\cite{mehboudi_mnflow_2024} to  design their DLD systems in an automated fashion. 
	In the following, we provide examples to explain this option.
\end{itemize}

By way of illustration, consider an application for which a DLD system with critical diameter of $d_c=30~\mu m$ is desired to isolate circulating tumor cell (CTC) clusters~\cite{au_microfluidic_2017}.
For a structure of periodicity of $N_p=10$,
one can use, for example, the \codex1{pow} model~\cite{ebadi_efficient_2019} 
to design the boundary profiles.
The DLD channel can be designed by using a single-command code as shown in Listing~\ref{lst_code_pow_wo_pressure_balance}.
There are some configurations that are set by the tool automatically in the absence of explicitly provided pertinent arguments.
In this example, it is assumed that pillars are circular and $g_w=g_a=d$, wherein $d$ denotes the pillars diameter, and is calculated to be about $64.7~\mu m$.
More details can be found in the original work~\cite{mehboudi_universal_2024} and the package documentation~\cite{mehboudi_mnflow_2024a}.
\noindent\begin{minipage}[!hb]{\linewidth}
\begin{lstlisting}[language=Python, label={lst_code_pow_wo_pressure_balance},
caption={
	A single-command code example usage of DDA~\cite{mehboudi_mnflow_2024} to design a DLD system and generate its mask layout CAD file, wherein $d_c=10~\mu m$, $N_p=10$, and boundary gap profiles are determined from the model developed by Ebadi \textit{et. al.}~\cite{ebadi_efficient_2019}.
}]
DLD(
	d_c=10,                          # Critical diameter (micron)
	Np=10,                           # Periodicity
	boundary_treatment='pow',        # Boundary profile
	opt_acc_balance_pressure=False,  # Disabling the pressure balance mechanism
)\end{lstlisting}
\end{minipage}

It is also feasible to add the pressure balance mechanism~\cite{inglis_fluidic_2020} when designing the boundary profiles by using a single-command code as shown in Listing~\ref{lst_code_pow_w_pressure_balance}.
It is worth mentioning that
the DDA is currently configured to apply the pressure balance mechanism by default. In order to disable this feature, one can pass  \codex1{opt\_acc\_balance\_pressure=False} as shown in Listing~\ref{lst_code_pow_wo_pressure_balance}, in which case 
the original model, \codex1{pow} in this example, will be used when designing the boundary profiles.
\noindent\begin{minipage}[!hb]{\linewidth}
\begin{lstlisting}[language=Python, label={lst_code_pow_w_pressure_balance},
caption={
	A single-command code example usage of DDA~\cite{mehboudi_mnflow_2024} to design a DLD system and generate its mask layout CAD file, wherein $d_c=10~\mu m$, $N_p=10$, and boundary gap profiles are determined from the model developed by Ebadi \textit{et. al.}~\cite{ebadi_efficient_2019} in conjunction with the pressure balance mechanism introduced by Inglis \textit{et. al.}~\cite{inglis_fluidic_2020}.
}]
DLD(
	d_c=10,                          # Critical diameter (micron)
	Np=10,                           # Periodicity
	boundary_treatment='pow',        # Boundary profile
)\end{lstlisting}
\end{minipage}

\begin{figure*}
	\centering
	\includegraphics[width=\textwidth]{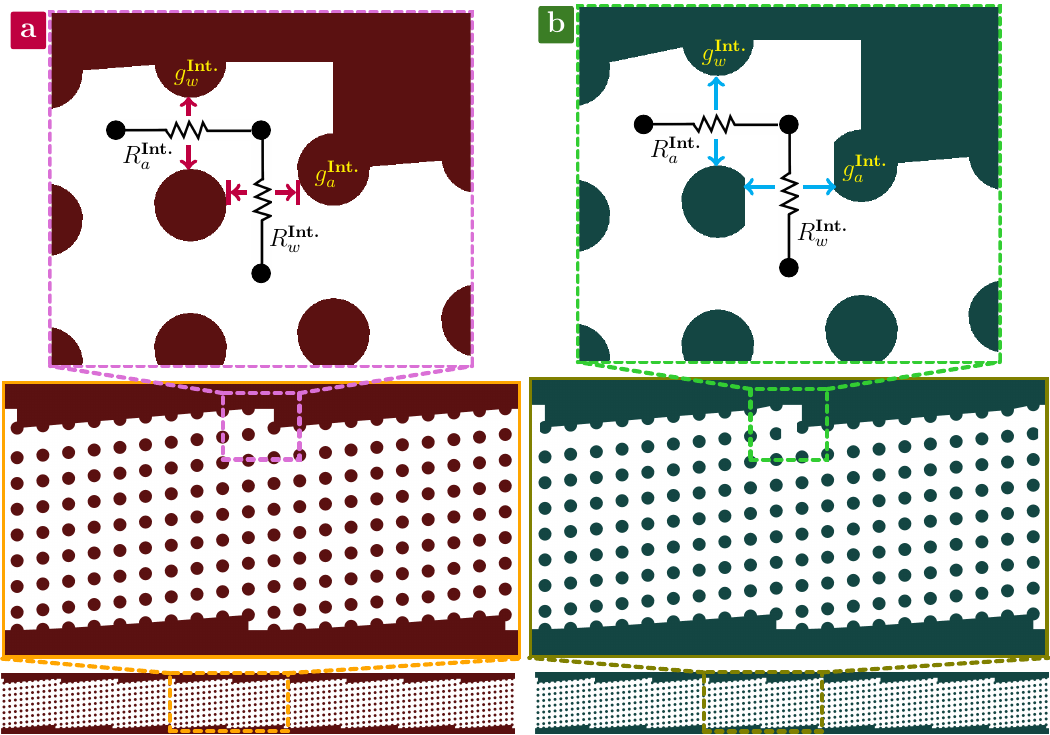}
	\caption{Schematic layout extracted from the CAD files 
	for DLD arrays with $N_p=10$ and $d_c=10~\mu m$, 
	wherein boundary gap profiles are determined by using the \codex1{pow} model~\cite{ebadi_efficient_2019}
	without (a) and with (b) pressure balance~\cite{inglis_fluidic_2020},
	designed and generated by 
	the DDA tool~\cite{mehboudi_mnflow_2024} 
	using the single-command codes shown in Listings~\ref{lst_code_pow_wo_pressure_balance} and \ref{lst_code_pow_w_pressure_balance}, respectively.
	The empirical formula provided by Davis~\cite{davis_microfluidic_2008} is used by the DDA tool to determine the configurations for bulk of the array to meet the desired critical diameter $d_c=30~\mu m$: $g_w=g_a\approx 64.7~\mu m$ and $\lambda_w=\lambda_a=129.4~\mu m$.
	}
	\label{fig_res_cad_pow}
\end{figure*}

The schematic CAD layouts generated by codes provided in Listings~\ref{lst_code_pow_wo_pressure_balance} and \ref{lst_code_pow_w_pressure_balance} 
are shown in Fig.~\ref{fig_res_cad_pow}.
The schematics can highlight the differences between boundary gap profiles with and without the pressure balance mechanism.
It can also be used to qualitatively explain the importance of the pressure balance concept as described in the remainder of this section.

Typically, a unit cell on accumulation sidewall has two outlets:
one being reserved for streamlines outlining the zizag mode, \textit{i.e.}, first stream split from fluid influx into the unit cell, which flows into the neighboring unit cell with a lower lane index,
and the other facilitating the transition of the rest of fluid influx to the downstream neighboring unit cell on the same fluidic lane, \textit{i.e.}, streamlines outlining the bump mode.

A key distinction of interface unit cell relative to others residing on accumulation sidewall is that it serves as a single-inlet, single-outlet conduit; it needs to pass all its fluid influx to its neighboring unit cell with a lower lane index.
In the case that the pressure balance mechanism is not present, Fig.~\ref{fig_res_cad_pow} (a), 
the net hydraulic resistance of interface unit cell, \textit{i.e.}, 
$R_a^\text{Int.}+R_w^\text{Int.}=2R_a$ (assuming $R_w=R_a$),
becomes significantly larger than that farther away from the interface unit cell, \textit{i.e.},
$R_a(1+\varepsilon)$,
which can cause a relatively large fraction of fluid (compared to $1/N_p$) from the upstream unit cells to flow through low-resistance paths 
by switching to lower neighboring fluidic lane (lower lane index)
leading to noticeably large values of local first stream width.
Regardless of how well the gap profile is designed on Rows \#0 to $N_p-2$,
an interface unit cell that is not configured properly on Row \#$N_p-1$
can still cause noticeable disturbances, 
which can adversely affect the fluid flow pattern even farther away from the cell,
deteriorating the separation performance of system as reported in Section~\ref{sec_res_gap_profile}.

%% file: conclusion.tex
\section{Conclusion}
\label{sec_conclusion}

In this work, we explored how pressure balance next to accumulation sidewall of DLD structure can impact fluid flow characteristics.
In particular we explored the uniformity of key variables  
such as first stream width and flux fraction
that play a significant role in particle separation efficiency.
We implemented the pressure balance mechanism in our DDA tool so that it can be integrated into different boundary treatment models which aim at providing optimal boundary gaps.
We studied various sidewall designs numerically with and without applying pressure balance scheme and investigated the effects of geometrical configurations of sidewalls on variations of critical diameter and first stream flux fraction across channel.
Our results show that all models studied in this work can benefit from the pressure balance mechanism noticeably.
In particular, the integration of \codex1{pow} model~\cite{ebadi_efficient_2019} and pressure balance model~\cite{inglis_fluidic_2020} showed the least nonuniformity of first stream width and flux fraction.
We also found that for a given boundary gap distribution,
there can be two desired zones for pressure balance parameter to minimize the nonuniformity of critical diameter. 
The first zone is associated with a substantial widening of the axial gap in adjacent pillars located in the interface unit cell near the accumulation sidewall. The second zone is linked to a substantial widening of the lateral gap between pillars in the interface unit cell.
We expect that this work will enable the design of DLD systems with improved separation performance.

%% file: end_notes.tex
\input{contribution}
\input{conflicts_of_interest}
\input{data_availability}

\input{acknowledgements}

%% file: contribution.tex
\section*{Author Contributions}
\textbf{Aryan Mehboudi:} Conceptualization, Methodology, Software, Validation, Formal analysis, Investigation, Resources, Data Curation, Writing - Original Draft, Writing - Review \& Editing, Visualization, Project administration. 
\textbf{Shrawan Singhal:}  
Resources, 
Writing - Review \& Editing, 
Supervision. 
\textbf{S.V. Sreenivasan:}
Resources, Supervision, Funding acquisition.

%% file: conflicts_of_interest.tex
\section*{Conflicts of interest}
There are no conflicts to declare.

%% file: data_availability.tex
\section*{Data Availability Statement}
The developed code for particle tracking is accessible from 
\href{https://github.com/am-0code1/fspt}{https://github.com/am-0code1/fspt}.
%
The code for DDA is part of {\mnflow} package and can be found at \href{https://github.com/am-0code1/mnflow}{https://github.com/am-0code1/mnflow} with DOI:
\href{https://doi.org/10.5281/zenodo.14357811}{10.5281/zenodo.14357811}. The version of the code employed for this study is version 
0.0.1.

%% file: acknowledgements.tex
\section*{Acknowledgements}
We greatly appreciate the anonymous reviewers for taking the time to review the manuscript and for their contribution in this work through providing 
helpful comments.

%% file: main.bbl
\providecommand{\href}[2]{#2}\begingroup\raggedright\begin{thebibliography}{10}

\bibitem{munaz_recent_2018}
A.~Munaz, M.~J.~A. Shiddiky, and N.-T. Nguyen, ``Recent advances and current
  challenges in magnetophoresis based micro magnetofluidics,''
  \href{http://dx.doi.org/10.1063/1.5035388}{{\em Biomicrofluidics} {\bfseries
  12} no.~3, (June, 2018) 031501}. \url{https://doi.org/10.1063/1.5035388}.

\bibitem{pethig_review_2010}
R.~Pethig, ``Review {{Article}}---{{Dielectrophoresis}}: {{Status}} of the
  theory, technology, and applications,''
  \href{http://dx.doi.org/10.1063/1.3456626}{{\em Biomicrofluidics} {\bfseries
  4} no.~2, (June, 2010) 022811}. \url{https://doi.org/10.1063/1.3456626}.

\bibitem{zhang_dep-on-a-chip_2019}
H.~Zhang, H.~Chang, and P.~Neuzil, ``{{DEP-on-a-Chip}}: {{Dielectrophoresis
  Applied}} to {{Microfluidic Platforms}},''
  \href{http://dx.doi.org/10.3390/mi10060423}{{\em Micromachines} {\bfseries
  10} no.~6, (June, 2019) 423}. \url{https://www.mdpi.com/2072-666X/10/6/423}.

\bibitem{wu_acoustofluidic_2019}
M.~Wu, A.~Ozcelik, J.~Rufo, Z.~Wang, R.~Fang, and T.~Jun~Huang,
  ``Acoustofluidic separation of cells and particles,''
  \href{http://dx.doi.org/10.1038/s41378-019-0064-3}{{\em Microsystems \&
  Nanoengineering} {\bfseries 5} no.~1, (June, 2019) 1--18}.
  \url{https://www.nature.com/articles/s41378-019-0064-3}.

\bibitem{zhou_viscoelastic_2020}
J.~Zhou and I.~Papautsky, ``Viscoelastic microfluidics: Progress and
  challenges,'' \href{http://dx.doi.org/10.1038/s41378-020-00218-x}{{\em
  Microsystems \& Nanoengineering} {\bfseries 6} no.~1, (Dec., 2020) 1--24}.
  \url{https://www.nature.com/articles/s41378-020-00218-x}.

\bibitem{carlo_inertial_2009}
D.~D. Carlo, ``Inertial microfluidics,''
  \href{http://dx.doi.org/10.1039/B912547G}{{\em Lab on a Chip} {\bfseries 9}
  no.~21, (Nov., 2009) 3038--3046}.
  \url{https://pubs.rsc.org/en/content/articlelanding/2009/lc/b912547g}.

\bibitem{zhang_fundamentals_2015}
J.~Zhang, S.~Yan, D.~Yuan, G.~Alici, N.-T. Nguyen, M.~E. Warkiani, and W.~Li,
  ``Fundamentals and applications of inertial microfluidics: A review,''
  \href{http://dx.doi.org/10.1039/C5LC01159K}{{\em Lab on a Chip} {\bfseries
  16} no.~1, (Dec., 2015) 10--34}.
  \url{https://pubs.rsc.org/en/content/articlelanding/2016/lc/c5lc01159k}.

\bibitem{ji_silicon-based_2008}
H.~M. Ji, V.~Samper, Y.~Chen, C.~K. Heng, T.~M. Lim, and L.~Yobas,
  ``Silicon-based microfilters for whole blood cell separation,''
  \href{http://dx.doi.org/10.1007/s10544-007-9131-x}{{\em Biomedical
  Microdevices} {\bfseries 10} no.~2, (Apr., 2008) 251--257}.
  \url{http://link.springer.com/10.1007/s10544-007-9131-x}.

\bibitem{huang_continuous_2004}
L.~R. Huang, E.~C. Cox, R.~H. Austin, and J.~C. Sturm, ``Continuous {{Particle
  Separation Through Deterministic Lateral Displacement}},''
  \href{http://dx.doi.org/10.1126/science.1094567}{{\em Science} {\bfseries
  304} no.~5673, (May, 2004) 987--990}.
  \url{https://www.science.org/doi/10.1126/science.1094567}.

\bibitem{yamada_pinched_2004}
M.~Yamada, M.~Nakashima, and M.~Seki, ``Pinched {{Flow Fractionation}}:
  {{Continuous Size Separation}} of {{Particles Utilizing}} a {{Laminar Flow
  Profile}} in a {{Pinched Microchannel}},''
  \href{http://dx.doi.org/10.1021/ac049863r}{{\em Analytical Chemistry}
  {\bfseries 76} no.~18, (Sept., 2004) 5465--5471}.
  \url{https://doi.org/10.1021/ac049863r}.

\bibitem{takagi_continuous_2005}
J.~Takagi, M.~Yamada, M.~Yasuda, and M.~Seki, ``Continuous particle separation
  in a microchannel having asymmetrically arranged multiple branches,''
  \href{http://dx.doi.org/10.1039/B501885D}{{\em Lab on a Chip} {\bfseries 5}
  no.~7, (June, 2005) 778--784}.
  \url{https://pubs.rsc.org/en/content/articlelanding/2005/lc/b501885d}.

\bibitem{yamada_hydrodynamic_2005}
M.~Yamada and M.~Seki, ``Hydrodynamic filtration for on-chip particle
  concentration and classification utilizing microfluidics,''
  \href{http://dx.doi.org/10.1039/b509386d}{{\em Lab on a Chip} {\bfseries 5}
  no.~11, (2005) 1233}. \url{http://xlink.rsc.org/?DOI=b509386d}.

\bibitem{yang_microfluidic_2006}
S.~Yang, A.~{\"U}ndar, and J.~D. Zahn, ``A microfluidic device for continuous,
  real time blood plasma separation,''
  \href{http://dx.doi.org/10.1039/B516401J}{{\em Lab on a Chip} {\bfseries 6}
  no.~7, (June, 2006) 871--880}.
  \url{https://pubs.rsc.org/en/content/articlelanding/2006/lc/b516401j}.

\bibitem{liang_scaling_2020}
W.~Liang, R.~H. Austin, and J.~C. Sturm, ``Scaling of deterministic lateral
  displacement devices to a single column of bumping obstacles,''
  \href{http://dx.doi.org/10.1039/D0LC00570C}{{\em Lab on a Chip} {\bfseries
  20} no.~18, (Sept., 2020) 3461--3467}.
  \url{https://pubs.rsc.org/en/content/articlelanding/2020/lc/d0lc00570c}.

\bibitem{wunsch_gel-on-a-chip_2019}
B.~H. Wunsch, S.-C. Kim, S.~M. Gifford, Y.~Astier, C.~Wang, R.~L. Bruce, J.~V.
  Patel, E.~A. Duch, S.~Dawes, G.~Stolovitzky, and J.~T. Smith,
  ``Gel-on-a-chip: Continuous, velocity-dependent {{DNA}} separation using
  nanoscale lateral displacement,''
  \href{http://dx.doi.org/10.1039/C8LC01408F}{{\em Lab on a Chip} {\bfseries
  19} no.~9, (Apr., 2019) 1567--1578}.
  \url{https://pubs.rsc.org/en/content/articlelanding/2019/lc/c8lc01408f}.

\bibitem{wunsch_nanoscale_2016}
B.~H. Wunsch, J.~T. Smith, S.~M. Gifford, C.~Wang, M.~Brink, R.~L. Bruce, R.~H.
  Austin, G.~Stolovitzky, and Y.~Astier, ``Nanoscale lateral displacement
  arrays for the separation of exosomes and colloids down to 20 nm,''
  \href{http://dx.doi.org/10.1038/nnano.2016.134}{{\em Nature Nanotechnology}
  {\bfseries 11} no.~11, (Nov., 2016) 936--940}.
  \url{http://www.nature.com/articles/nnano.2016.134}.

\bibitem{smith_integrated_2018}
J.~T. Smith, B.~H. Wunsch, N.~Dogra, M.~E. Ahsen, K.~Lee, K.~K. Yadav, R.~Weil,
  M.~A. Pereira, J.~V. Patel, E.~A. Duch, J.~M. Papalia, M.~F. Lofaro,
  M.~Gupta, A.~K. Tewari, C.~{Cordon-Cardo}, G.~Stolovitzky, and S.~M. Gifford,
  ``Integrated nanoscale deterministic lateral displacement arrays for
  separation of extracellular vesicles from clinically-relevant volumes of
  biological samples,'' \href{http://dx.doi.org/10.1039/C8LC01017J}{{\em Lab on
  a Chip} {\bfseries 18} no.~24, (Dec., 2018) 3913--3925}.
  \url{http://pubs.rsc.org/en/content/articlelanding/2018/lc/c8lc01017j}.

\bibitem{davis_deterministic_2006}
J.~A. Davis, D.~W. Inglis, K.~J. Morton, D.~A. Lawrence, L.~R. Huang, S.~Y.
  Chou, J.~C. Sturm, and R.~H. Austin, ``Deterministic hydrodynamics:
  {{Taking}} blood apart,''
  \href{http://dx.doi.org/10.1073/pnas.0605967103}{{\em Proceedings of the
  National Academy of Sciences} {\bfseries 103} no.~40, (Oct., 2006)
  14779--14784}.
  \url{https://www-pnas-org.ezproxy.lib.utexas.edu/doi/full/10.1073/pnas.0605967103}.

\bibitem{salafi_review_2019}
T.~Salafi, Y.~Zhang, and Y.~Zhang, ``A {{Review}} on {{Deterministic Lateral
  Displacement}} for {{Particle Separation}} and {{Detection}},''
  \href{http://dx.doi.org/10.1007/s40820-019-0308-7}{{\em Nano-Micro Letters}
  {\bfseries 11} no.~1, (Sept., 2019) 77}.
  \url{https://doi.org/10.1007/s40820-019-0308-7}.

\bibitem{inglis_critical_2006}
D.~W. Inglis, J.~A. Davis, R.~H. Austin, and J.~C. Sturm, ``Critical particle
  size for fractionation by deterministic lateral displacement,''
  \href{http://dx.doi.org/10.1039/B515371A}{{\em Lab on a Chip} {\bfseries 6}
  no.~5, (May, 2006) 655--658}.
  \url{https://pubs.rsc.org/en/content/articlelanding/2006/lc/b515371a}.

\bibitem{davis_microfluidic_2008}
J.~A. Davis, {\em Microfluidic {{Separation}} of {{Blood Components}} through
  {{Deterministic Lateral Displacement}}}.
\newblock PhD thesis, Princeton University, 2008.
\newblock \url{https://swh.princeton.edu/~sturmlab/theses/Davis-Thesis.pdf}.

\bibitem{inglis_efficient_2009}
D.~W. Inglis, ``Efficient microfluidic particle separation arrays,''
  \href{http://dx.doi.org/10.1063/1.3068750}{{\em Applied Physics Letters}
  {\bfseries 94} no.~1, (Jan., 2009) 013510}.
  \url{https://doi.org/10.1063/1.3068750}.

\bibitem{pariset_anticipating_2017}
E.~Pariset, C.~Pudda, F.~Boizot, N.~Verplanck, J.~Berthier, A.~Thuaire, and
  V.~Agache, ``Anticipating {{Cutoff Diameters}} in {{Deterministic Lateral
  Displacement}} ({{DLD}}) {{Microfluidic Devices}} for an {{Optimized Particle
  Separation}},'' \href{http://dx.doi.org/10.1002/smll.201701901}{{\em Small}
  {\bfseries 13} no.~37, (2017) 1701901}.
  \url{https://onlinelibrary.wiley.com/doi/abs/10.1002/smll.201701901}.

\bibitem{feng_maximizing_2017}
S.~Feng, A.~M. Skelley, A.~G. Anwer, G.~Liu, and D.~W. Inglis, ``Maximizing
  particle concentration in deterministic lateral displacement arrays,''
  \href{http://dx.doi.org/10.1063/1.4981014}{{\em Biomicrofluidics} {\bfseries
  11} no.~2, (Mar., 2017) 024121}.
  \url{https://aip.scitation.org/doi/10.1063/1.4981014}.

\bibitem{ebadi_efficient_2019}
A.~Ebadi, M.~J. Farshchi~Heydari, R.~Toutouni, B.~Chaichypour, M.~Fathipour,
  and K.~Jafari, ``Efficient paradigm to enhance particle separation in
  deterministic lateral displacement arrays,''
  \href{http://dx.doi.org/10.1007/s42452-019-1064-5}{{\em SN Applied Sciences}
  {\bfseries 1} no.~10, (Sept., 2019) 1184}.
  \url{https://doi.org/10.1007/s42452-019-1064-5}.

\bibitem{inglis_fluidic_2020}
D.~Inglis, R.~Vernekar, T.~Kr{\"u}ger, and S.~Feng, ``The fluidic resistance of
  an array of obstacles and a method for improving boundaries in deterministic
  lateral displacement arrays,''
  \href{http://dx.doi.org/10.1007/s10404-020-2323-x}{{\em Microfluidics and
  Nanofluidics} {\bfseries 24} no.~3, (Feb., 2020) 18}.
  \url{https://doi.org/10.1007/s10404-020-2323-x}.

\bibitem{mehboudi_universal_2024}
A.~Mehboudi, S.~Singhal, and S.~Sreenivasan, ``A universal framework for design
  and manufacture of deterministic lateral displacement chips,''
  \href{http://dx.doi.org/10.1039/D4LC00838C}{{\em Lab on a Chip} (Dec., 2024)
  }. \url{https://pubs.rsc.org/en/content/articlelanding/2025/lc/d4lc00838c}.

\bibitem{mehboudi_mnflow_2024}
A.~Mehboudi, ``{{mnFlow}}: {{A}} package for micro/nanoflow,'' Dec., 2024.
\newblock \url{https://zenodo.org/doi/10.5281/zenodo.14357811}.

\bibitem{au_microfluidic_2017}
S.~H. Au, J.~Edd, A.~E. Stoddard, K.~H.~K. Wong, F.~Fachin, S.~Maheswaran,
  D.~A. Haber, S.~L. Stott, R.~Kapur, and M.~Toner, ``Microfluidic
  {{Isolation}} of {{Circulating Tumor Cell Clusters}} by {{Size}} and
  {{Asymmetry}},'' \href{http://dx.doi.org/10.1038/s41598-017-01150-3}{{\em
  Scientific Reports} {\bfseries 7} no.~1, (May, 2017) 2433}.
  \url{https://www.nature.com/articles/s41598-017-01150-3}.

\bibitem{mehboudi_mnflow_2024a}
A.~Mehboudi, ``{{mnFlow Documentation}},'' 2024.
\newblock \url{https://mnflow.readthedocs.io/en/latest/}.

\end{thebibliography}\endgroup
